\documentclass[usenatbib]{mn2e}
\usepackage{graphicx}

\begin{document}

\title[Bow shocks from runaway early type stars]{Modelling multi-wavelength 
  observational characteristics of bow shocks from runaway early type stars}
\author[David M. Acreman, Ian R. Stevens and Tim J. Harries]{David
  M. Acreman$^1$\thanks{ E-mail acreman@astro.ex.ac.uk}, Ian R. Stevens$^2$ and Tim J. Harries$^1$ \\
$^1$ School of Physics and Astronomy, University of Exeter, Stocker Road, Exeter EX4
4QL.\\
$^2$School of Physics and Astronomy, University of Birmingham,
Edgbaston, Birmingham, B15 2TT, UK}

\maketitle

\begin{abstract}

  We assess the multi-wavelength observable properties of the bow shock around a
  runaway early type star using a combination of hydrodynamical
  modelling, radiative transfer calculations and synthetic imaging.
  Instabilities associated with the forward shock produce dense knots of material
  which are warm, ionised and contain dust. These knots of material are responsible for 
  the majority of emission at far infra-red, H$\alpha$ and
  radio wavelengths. The large scale bow shock morphology
  is very similar and differences are primarily due to variations in the assumed spatial resolution. 
  However infra-red intensity slices (at 22~microns and 12~microns) show that the effects 
  of a temperature gradient can be resolved at a realistic spatial resolution for an object 
  at a distance of 1~kpc. 

\end{abstract}


\section{Introduction}

Stars with large velocities ($>40\rm{km~s}^{-1}$) relative to the surrounding interstellar
medium (ISM) are termed runaway stars
\citep{Blaauw1961,Cruz1974,Tetzlaff_2011}. 
Space velocities up to $200~\rm{km~s}^{-1}$ were observed by \cite{Blaauw1961} and 
if conditions in the ISM are favourable the velocity of a runaway star relative to the ISM will
be supersonic \citep{Huthoff2002} leading to the formation of a
bow shock \citep{Baranov1976}. Runaway early-type stars have strong
stellar winds and experience a particularly strong interaction with the
ISM, due to the high space velocity of the star and the high
velocity of the wind. A number of high velocity early-type stars
are were identified by \cite{Blaauw1961} and \cite{Cruz1974}. Bow shocks have also seen around 
evolved stars, e.g. red super-giants \citep{Noriega-Crespo1997,cox_2012,Gvaramadze2014,Meyer2014b}. These bow 
shocks are similar to those around early-type, main sequence stars albeit with a 
lower stellar wind velocity. 

It has been proposed that runaway stars are produced by binary
supernovae \citep{Blaauw1961} or result from dynamical interactions
between stars in a cluster \citep{Poveda1967}. There is evidence that,
in practice, runaway stars are produced by both mechanisms
\citep{PortegiesZwart2000,Hoogerwerf_2000,Hoogerwerf2001}. Runaway
stars with velocities of a few 100~km~s$^{-1}$ can be generated by both
the binary supernova \citep{PortegiesZwart2000} and dynamical ejection
scenarios \citep{Poveda1967,Leonard1991}. 
Moreover, ejection from the Galactic centre, due to interaction with
the supermassive black hole, has been cited as a mechanism for the
formation of hypervelocity stars with even higher spatial velocities
\citep{Hills1988,Brown_2012}. In addition to bow shocks around stars
with large peculiar velocities, young stars in a cluster can drive
outflows, which cause ISM velocities to deviate from the local
standard of rest, and bow shocks can be generated without ejecting
stars from the cluster \citep{Povich_2008}.

The formation of isolated massive stars is an important factor in
discriminating between different star formation mechanisms. In the
competitive accretion paradigm massive stars form by accreting
material in a cluster environment
\citep{Zinnecker1982,Larson1992,Bonnell1997,Bonnell2004} whereas in
the core accretion paradigm massive stars are able to form in
isolation \citep{Wolfire1987,Yorke2002,Krumholz2009}. When searching
for evidence of isolated massive star formation runaways need to be
excluded, as they are likely to be found far from where they formed
\citep{deWit2005,Bressert2012}. Runaway stars can sometimes be
identified through proper motion observations or line-of-sight
velocity measurements but the presence of a bow shock is also a
valuable indication that a star is a runaway. Stellar bow shocks yield
not only kinematic information but the stand-off distance of the shock
has been used as a diagnostic of stellar wind and ISM properties
\citep{Brown_2005,Povich_2008,Kobulnicky_2010,cox_2012}. As well as
using bow shocks as diagnostics it is important to understand the
observable properties of stellar bow shocks, as they may produce an
infra-red excess which could be confused with emission from a
circumstellar disc \citep{Povich_2008,Gaspar_2008}.

Observational studies have detected stellar bow shocks in various
regions of the spectrum, most notably in the mid-infrared
\citep{NC97,vanBuren_1988,Gaspar_2008,Povich_2008,Kobulnicky_2010,Peri_2012}
but also in other regimes e.g. H$\alpha$ observations by
\cite{Brown_2005}. A multi-wavelength study of the O5 star HD~192281 by
\cite{Arnal_2011} revealed an H{\sc{i}} structure consistent with
formation by a bow shock, with counterparts seen in infrared, radio
continuum and CO (1--0) observations. The radio emission had a thermal
spectrum with ionisation as the proposed source. Non-thermal radio emission,
from relativistic particles, has been detected in observations
of BD+43$^{\circ}$3654 by \cite{Benaglia_2010} and non-thermal X-ray emission 
from the AE Aur bow shock was detected by \cite{Lopez2012}. 

Analytical expressions can be derived for some bow shock properties, given certain 
simplifying assumptions. \cite{Wilkin_1996} treats the bow shock as an 
axisymmetric thin shell and calculates analytical expressions for the shape, mass 
column density and velocity of the shocked gas. However when the simplifying assumptions 
are relaxed numerical modelling is required. In their study of ultracompact H{\sc{ii}} regions
\cite{MacLow1991} use a numerical model of a shell in momentum balance. This gives a two-dimensional 
solution which they use to produce synthetic radio observations. \cite{Raga1997} use two-dimensional 
models, with hydrodynamics and a radiative transfer approximation, to produce synthetic dust and 
H$\alpha$ emission maps for a runaway O star. 
A number of subsequent studies have also made use of two-dimensional hydrodynamics to study the physics of bow shocks 
around runaway massive stars e.g. \cite{Comeron1998,cox_2012}.
More recently \citep{Meyer2014} modelled an evolving 
main sequence star, using a two-dimensional hydrodynamics code, and generated synthetic H$\alpha$ observations.

In this paper we model the bow shock caused by a runaway early type
star and determine the observational properties of the
resulting structure. To achieve this we combine hydrodynamical
modelling (described in Section~\ref{section:vh1_hydro}) with
Monte-Carlo radiative transfer calculations (described in
Section~\ref{section:torus_photoion}) to model the radiation
influenced region around the star. We generate synthetic observables,
as described in Section~\ref{section:synthetic_obs}, focussing on the
infra-red, H$\alpha$ and radio regimes where successful detections of
stellar bow shocks have already been made. We discuss our results in
Section~\ref{section:discussion}, where we compare them to 
synthetic observations in the literature, and discuss assumptions and potential 
improvements to the model. We close with our conclusions in Section~\ref{section:conclusions}.

\section{Hydrodynamic calculation}
\label{section:vh1_hydro}

The formation of the bow shock structure was modelled using the VH-1
hydrodynamics code \citep{Blondin2003,VH1}, a grid-based code which uses the
Piecewise-Parabolic Method \citep{colella_1984} and provides good
resolution of shocks in complex flows. The gas is treated as an 
ideal gas with radiative cooling implemented as a cooling curve. 
The radiative cooling in the hydrodynamic calculation is modelled using
an analytic fit to the results of \cite{Schure2009}, assuming a solar
metallicity plasma. Radiation pressure is not included and no additional physical viscosity is added.

The simulation grid uses two dimensional, cylindrical polar co-ordinates, with the direction of
motion of the star aligned with the z-axis. 
The grid size is $10^{18}~\rm{cm}$ in each linear dimension with 1000 grid cells. 
The star is placed on the z-axis and ISM material flows onto the grid at a fixed velocity
of 100~km~s$^{-1}$ (i.e. the simulation is performed in the frame of
reference of the star). The star has a wind which causes material to
flow radially outwards and meet the incoming ISM in the upstream
direction. The stellar wind is implemented as a region 
with a constant outflow velocity which is imposed 
within a fixed radius of 50 grid cells from the star. 
The density in the outflow region decreases with radius such that the mass loss rate is constant.
The wind radius must be large 
enough to ensure that the wind remains spherical but needs to be smaller than 
the minimum extent of the bow shock region. The main model parameters are listed in
Table~\ref{tab:hydro_sim_params}.
\begin{table}
  \centering
  \caption{Parameters of the hydrodynamical simulation}
  \begin{tabular}{lll}
    \hline
    Parameter & & Value \\
    \hline
    Stellar mass-loss rate & $\dot{M}_W$ & $10^{-5} M_{\odot} \rm{yr}^{-1}$ \\
    Stellar wind velocity & $V_W$ & 2000~km~s$^{-1}$ \\
    Stellar velocity & $V_*$ & 100~km~s$^{-1}$ \\
    ISM number density & $n_{\rm{ISM}}$& 1000~$\rm{cm}^{-3}$ \\
    Grid size & & $10^{18}~\rm{cm}$ \\ 
    	Number of grid cells & & $1000^2$ \\
    \hline
  \end{tabular}
  \label{tab:hydro_sim_params}
\end{table}
Dust mass is not included in the hydrodynamical simulations, where its contribution to the mass is negligible. However radiation pressure can have a significant impact on dust dynamics which could influence the gas if the dust and gas are dynamically coupled \citep{villaver_2012,Ochsendorf2014,Gratier2014}. Dust opacity is included in the radiative transfer calculation, where the effect is significant (see Section~\ref{section:torus_photoion}).

\begin{figure*}
\begin{center}
\includegraphics[scale=0.475]{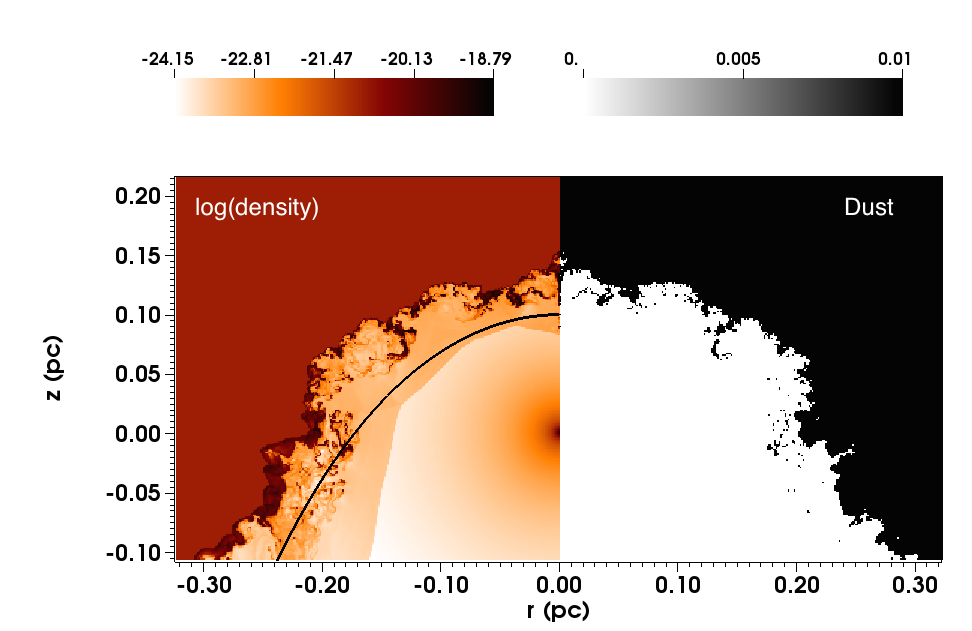}
\includegraphics[scale=0.475]{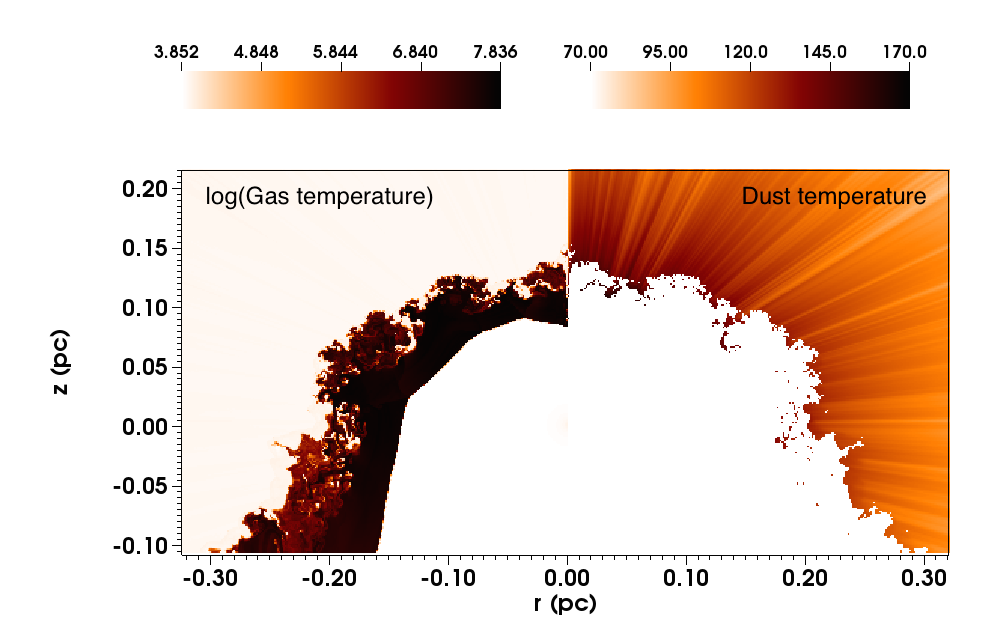}
\caption{Density (top left) in $\rm{gm~cm^{-3}}$, dust distribution (top right), gas temperature (bottom left) in K, and dust temperature (bottom right) in K, at the end of the photoionisation equilibrium calculation. The star is located at the origin of the grid. 
The black curve shown on the density plot is the Wilkin analytical solution. The dust distribution plot shows dust fraction by mass and is black where dust is present and white where there is no dust.}
\label{fig:torus_photoion_calc}
\end{center}
\end{figure*}

The density distribution is plotted in Fig.~\ref{fig:torus_photoion_calc} (top left) and
shows a two shock structure with a forward shock, where the ISM
shocks, and an inner shock where the stellar wind terminates. This is
similar to the structures seen in other simulations\citep[e.g.][]{Comeron1998,cox_2012,Mackey_2012,Meyer2014} where the two shocks
bound a contact discontinuity. However our model does not show a clear separation between the 
forward shock and contact discontinuity. If the shocked ambient gas cools very rapidly, as expected due to the high ISM 
density, then the outer layer of hot material is not present, as described by \cite{Comeron1998}. 
The forward shock is subject to instabilities which result in the 
formation of denser knots of material. Similar instabilities have been seen in other numerical models of stellar bow shocks (e.g. \cite{Comeron1998}, \cite{cox_2012} and \cite{Meyer2014}). A number of different mechanisms can generate these instabilities (e.g. Kelvin-Helmholtz instabilities, non-linear thin shell instabilities \citep{Vishniac_1994} and transverse acceleration instabilities \citep{Dgani_1996}) as discussed by \cite{Comeron1998}.

An analytical expression for the shock stand-off distance $R_0$ can be
derived by determining the distance at which the stellar wind ram
pressure and the ISM ram pressure are equal \citep{Wilkin_1996} 
\begin{equation}
R_0 = \sqrt{ \frac{\dot{M}_W V_W}{4 \pi \rho_{\rm{ISM}} V_*^2} }
\end{equation}
where $\rho_{\rm{ISM}}$ is the ISM mass density and other symbols are
defined in Table~\ref{tab:hydro_sim_params}.
We assume the ISM is composed of
hydrogen and helium with a 10:1 ratio and is fully ionised. The ISM number density $n_{\rm{ISM}}$ is the 
sum of the hydrogen ion, helium ion and electron number densities (i.e. $n_{\rm{ISM}}=n_{H^+}+n_{He^{++}}+n_{e^{-}}$)
and $\rho_{\rm{ISM}}=\frac{14}{23}n_{\rm{ISM}} m_H$ where $m_H$ is the mass of a
hydrogen atom. For the parameters used in this simulation the calculated stand-off distance is $R_0=0.10$~pc. \cite{Wilkin_1996} derives an 
analytical expression for the shape of the bow shock
\begin{equation}
R = R_0 \csc\theta \sqrt{3\left(1 - \theta \cot\theta \right)}
\end{equation}
where $R$ is the distance from the star and $\theta$ is the polar angle from the axis of symmetry. The analytical solution is shown in Fig.~\ref{fig:torus_photoion_calc} as a black curve overlaid on the density plot. In our model the z-axis location of the stand-off distance is between the forward shock and the wind termination (inner) shock. The shape of the forward shock follows the analytical 
solution but is located further from the star. The post-stellar wind material in our model is at a high temperature and the density ratio is 
approximately 4 which suggest an adiabatic shock with inefficient cooling of the shocked gas. \cite{Comeron1998} note that if cooling of the shocked stellar wind is inefficient then a layer of hot, low density gas lies between the stellar wind and the bow shock, causing the bow shock to move forward of $R_0$.

\section{Photoionisation calculation}
\label{section:torus_photoion}

The density distribution from the hydrodynamical calculation,
described in Section~\ref{section:vh1_hydro}, is used as input to the
{\sc{torus}}\footnote{www.astro.ex.ac.uk/people/th2/torus\_html/homepage.html}
radiative transfer code \citep{Harries00}.
The VH-1 density field is mapped onto a 
{\sc{torus}} grid comprising $256\times256$ cells\footnote{the number
  of cells in each dimension of the {\sc{torus}} grid must be a power
  of 2 as {\sc{torus}} uses a tree structure to store the grid}. This
density field is used to calculate temperature and ionisation
equilibrium using an iterative Monte-Carlo method, similar to that of
\cite{Ercolano_2003} and \cite{wood_2004}. The photoionisation
component of {\sc{torus}} has been described in detail by
\cite{Haworth2012a} and \cite{Haworth2015}. For the calculations presented in this paper the
photoionisation calculation operates stand-alone (without a
{\sc{torus}} hydrodynamics calculation) using a static density
distribution derived from the VH-1 calculation. Consequently we are
able to use a more computationally expensive determination of
equilibrium temperature (based on the balance between ionisation,
recombination and forbidden line cooling) rather than the simplified
thermal balance calculation of \cite{Haworth2012a}. Like
\cite{Haworth2012a} we use a polychromatic radiation field and
explicitly represent the diffuse radiation field (i.e. there is no
on-the-spot approximation). A total of 22 species are included in the
photoionisation calculation, with abundances from \cite{ferland_1995}
as shown in Table~\ref{tab:abundances}.
\begin{table}
  \centering
  \caption{Elements and species included in the photoionisation
    calculation with elemental abundances relative to hydrogen.}
  \begin{tabular}{lll}
    \hline
    Element & Species & Abundance \\
    \hline
   H & I,II          & 1 \\
   He & I,II,III    & 0.1 \\
   C & I,II,III,IV  & $2.2\times10^{-4}$ \\
   N & I,II,III      & $4\times10^{-5}$ \\
   O & I,II,III      & $3.3\times10^{-4}$ \\
   Ne & I,II,III    & $5\times10^{-5}$ \\
   S & I,II,III,IV   & $9\times10^{-6}$ \\
    \hline
  \end{tabular}
  \label{tab:abundances}
\end{table}
The star is represented by a blackbody source with a temperature of
40~000K and a luminosity of $L=1.41\times10^{6} L_{\odot}$ which produces an ionising photon flux of $7.45\times10^{49}~\rm{s}^{-1}$.

\begin{figure}
  \includegraphics[scale=0.3,angle=0]{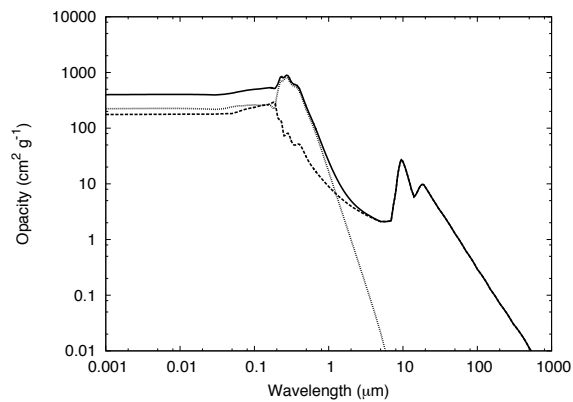}
  \caption{Total opacity (solid line), absorption opacity (dashed line) and scattering opacity (dotted line) per gramme of gas, 
  plotted as a function of wavelength, for the $0.1~\mu\rm{m}$ silicate dust grains used in the radiative transfer calculations.}
\label{fig:opacity}
\end{figure}

The effects of dust opacity are included in the radiative transfer calculation, and separate gas and dust temperatures are calculated as the gas and dust are not expected to be thermally coupled.
The dust grains are Draine and Lee silicates with a uniform size of $0.1~\mu\rm{m}$ and a density of 3.6~$\rm{g~cm^{-3}}$. The total opacity (solid line), absorption opacity 
(dashed line) and scattering opacity (dotted line) per gramme of gas are plotted as a function of wavelength in Fig.~\ref{fig:opacity} for the assumed dust to gas mass ratio of 0.01.
A prescribed dust distribution is added to the hydrodynamics 
results (which do not include dust) as a post-processing step. The 
dust distribution is shown in Fig.~\ref{fig:torus_photoion_calc} (top right).
The free-flowing stellar wind (defined as any cell with a velocity greater than 1800~km~s$^{-1}$) and the 
shocked wind (defined as any cell with a temperature greater than $1.1\times10^4$K) are dust free, and dust 
is present elsewhere.

There is substantial shock heating in the region between the forward shock and the stellar wind termination shock which is absent in a pure radiative transfer calculation. To account for this effect the temperature used by {\sc{torus}} in shocked stellar wind region (the region where the {\sc{VH-1}} temperature is greater than $1.1\times10^4$K) is constrained to be at least as great as the temperature calculated by the VH-1 hydrodynamical calculation. The gas and dust temperatures calculated by {\sc{torus}} are shown in the lower two panels of Fig.~\ref{fig:torus_photoion_calc}. The gas temperature in the inter-shock region is high ($\sim10^7\rm{K}$) and in this part of the domain the temperature is governed by the results of the hydrodynamical calculation. In this inter-shock region the temperature will be determined by the balance between shock heating and radiative cooling, and both these processes are included in the VH-1 hydrodynamical calculation. Elsewhere the gas temperature is determined by the radiative transfer calculation and the lower temperatures (approximately 7000-9000K) are close to the corresponding temperature assumed in the hydrodynamical calculation ($10^4\rm{K}$). All the gas in the simulation domain is highly ionised, consistent with the assumption of fully ionised gas in the hydrodynamical calculation. The dust temperature is much lower than the gas temperature and a ``sunburst'' pattern is seen in the dust temperature where dense knots in forward shock shadow the upstream dust. Similar features were seen in the planetary nebula simulations of \cite{Toala2014}.

\section{Synthetic observations}
\label{section:synthetic_obs}

The results from the photoionisation calculation were used to generate
synthetic observations (infrared, H$\alpha$ and radio images and
infrared SEDs). The temperature, density, abundance and dust
distributions at the end of the photoionisation calculation are used
to populate the {\sc{torus}} grid with emissivities at the appropriate
wavelength. A Monte-Carlo sampling technique \citep{Harries00} is then
used to sample the emission with a number of photon packets which are
propagated through the grid (accounting for scattering and absorption
events) to determine the flux in a prescribed image plane. The method
has previously been applied to output from {\sc{torus}}
photoionisation calculations by \cite{Haworth2012b}.

\subsection{Infrared images and SEDs}
\label{section:ir_images}

For generating infrared images and SEDs the emissivity of a grid cell
$\epsilon_{\lambda}$ at a wavelength $\lambda$ is given by
\begin{equation}
\epsilon_{\lambda} = B_{\lambda}\left(T\right) \kappa_{\lambda} \rho_{\rm{ISM}}
\end{equation}
where $B_{\lambda}\left(T\right)$ is the Planck function at
temperature $T$, $\kappa_{\lambda}$ is the wavelength dependent
absorption opacity from dust, and $\rho_{\rm{ISM}}$ is the ISM gas density. 
Referring to
Fig.~\ref{fig:torus_photoion_calc} we see that the emissivity
will be zero in the stellar wind (both shocked and un-shocked regions), 
as there is no dust and hence
$\kappa_{\lambda}=0$. The highest emissivities will be in the high
density parts of the forward shock where the density 
and temperature (hence $B_{\lambda}\left(T\right)$) are high.

\subsubsection{Images}

\begin{figure*}
  \centering
\includegraphics[scale=0.35,angle=0]{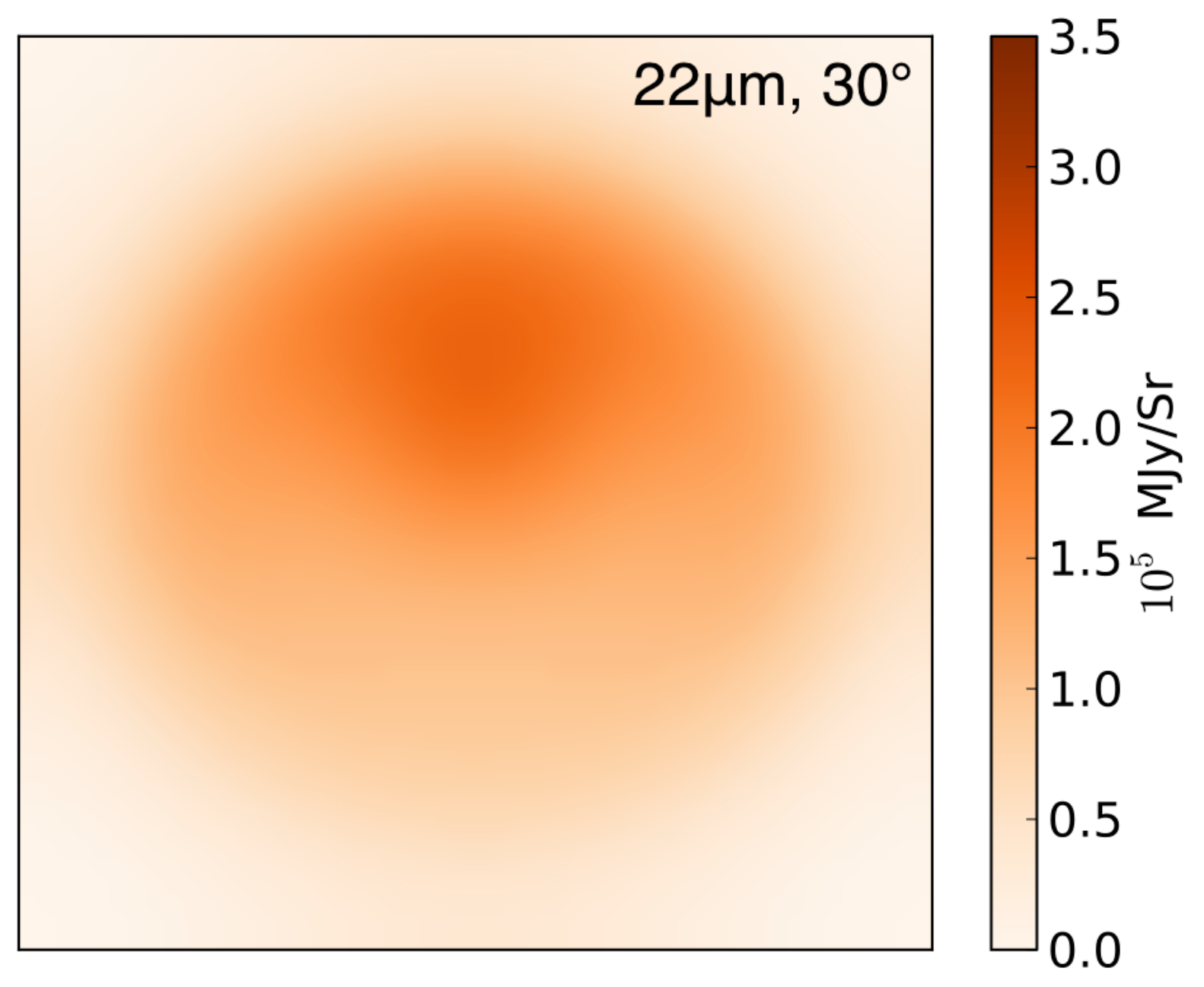}
\includegraphics[scale=0.35,angle=0]{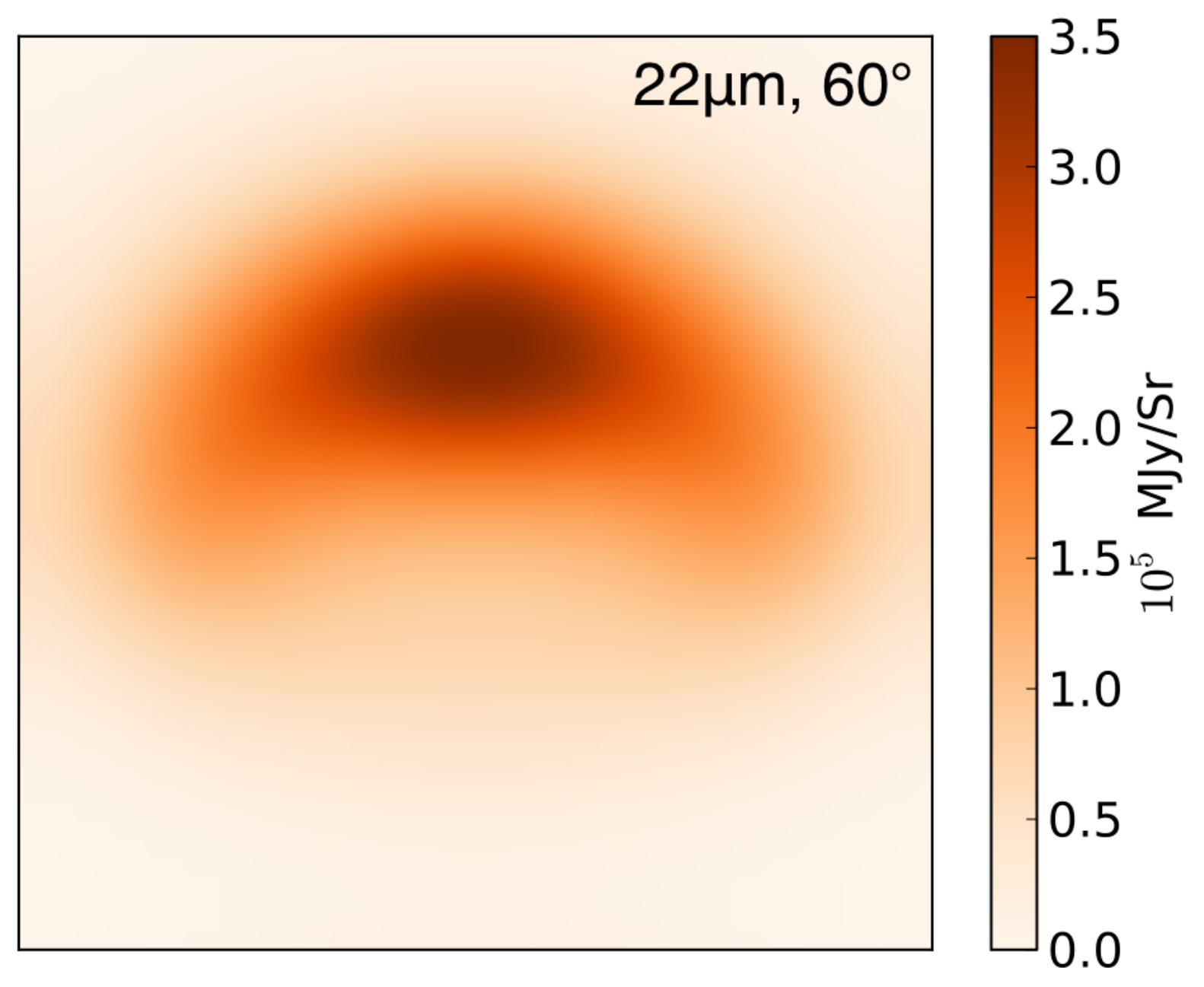}
\includegraphics[scale=0.35,angle=0]{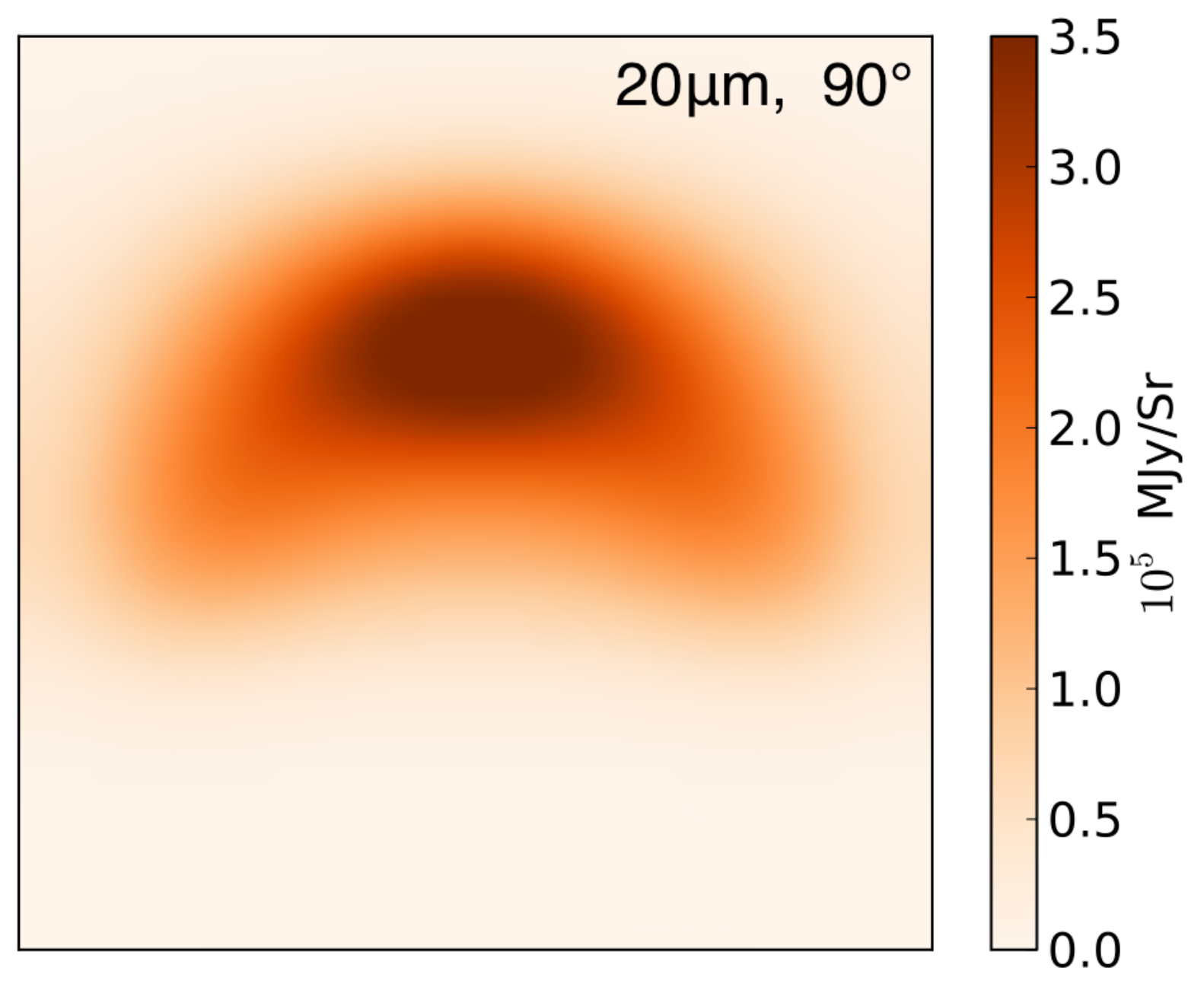}
\includegraphics[scale=0.35,angle=0]{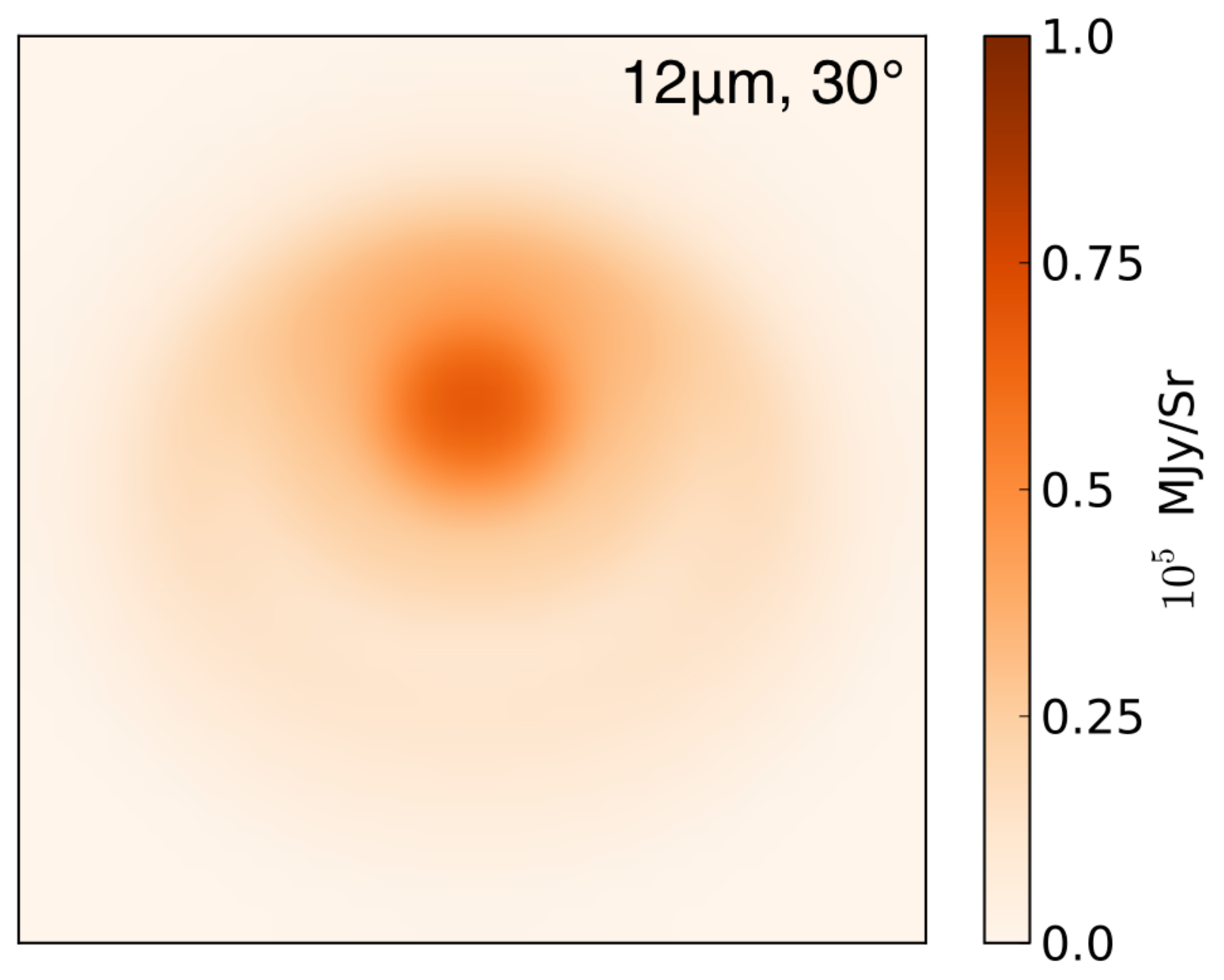}
\includegraphics[scale=0.35,angle=0]{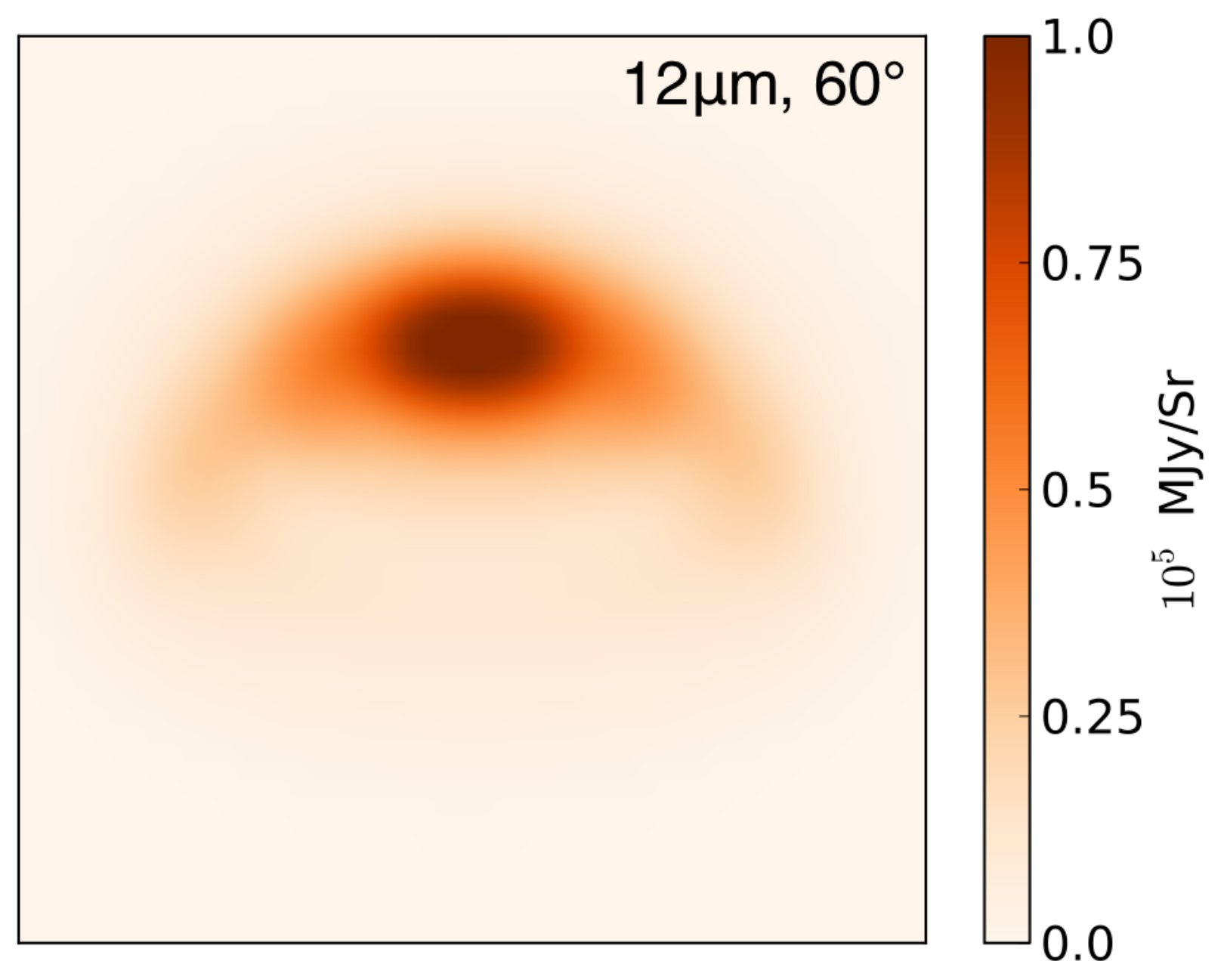}
\includegraphics[scale=0.35,angle=0]{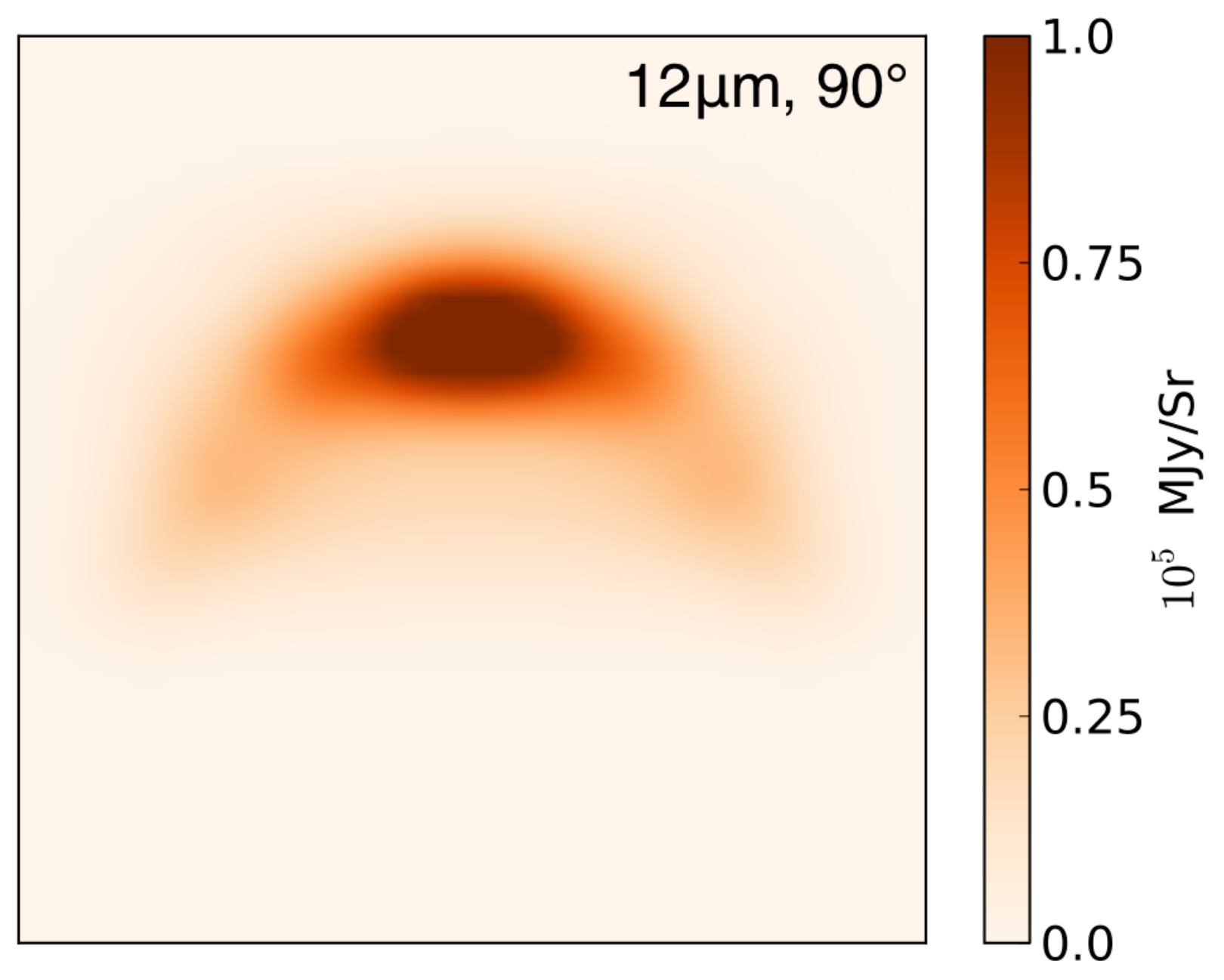}
\includegraphics[scale=0.35, angle=0]{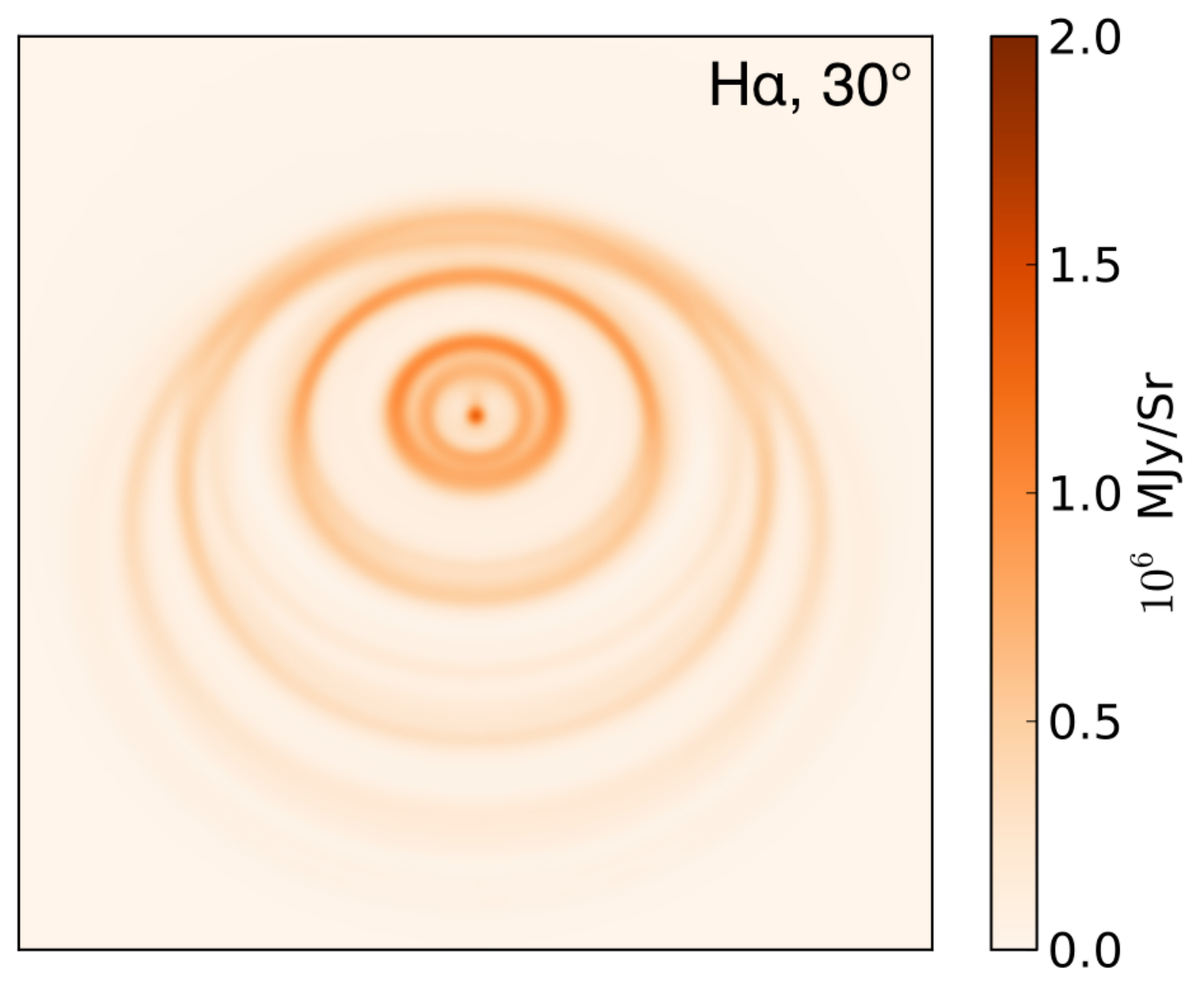}
\includegraphics[scale=0.35, angle=0]{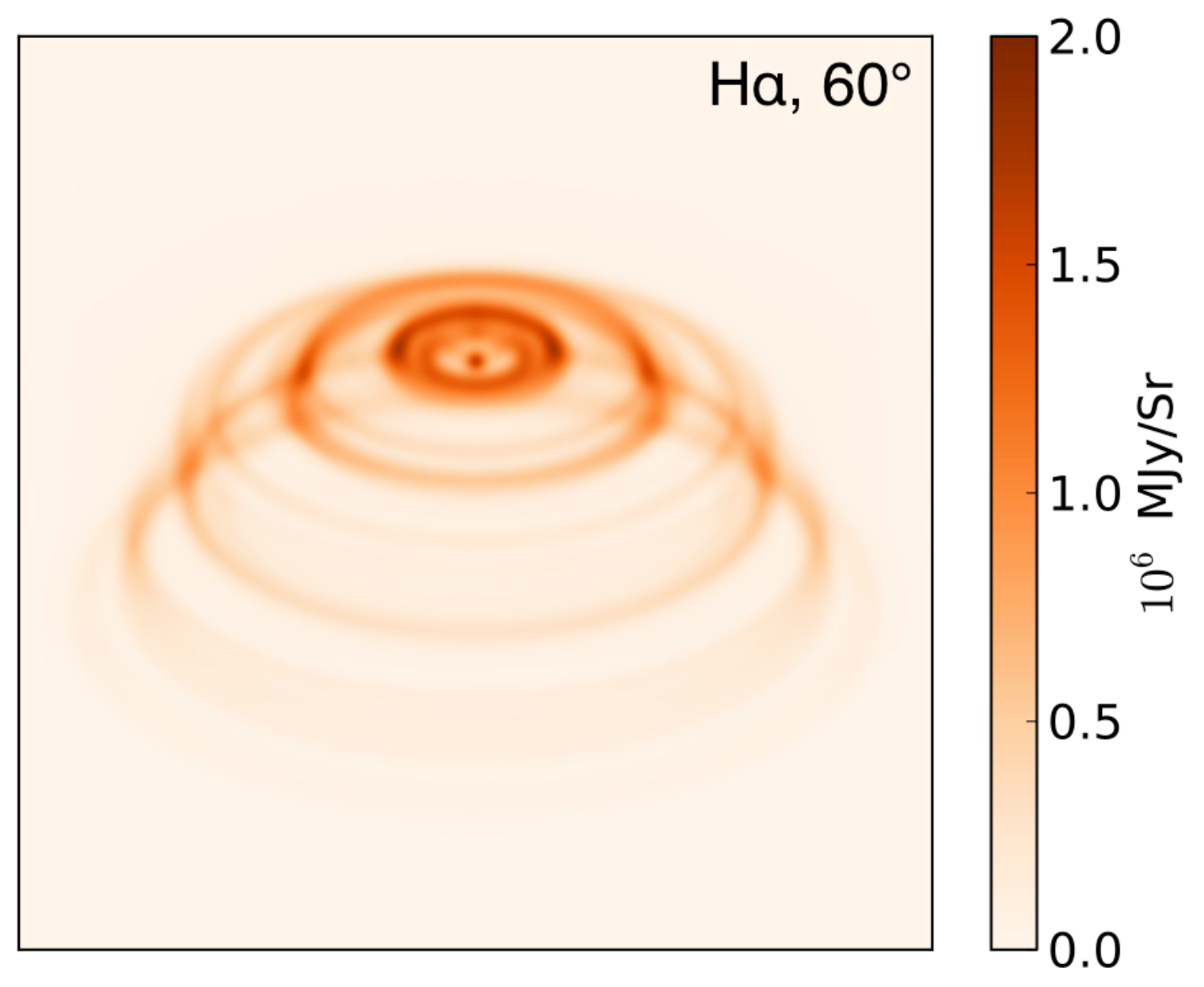}
\includegraphics[scale=0.35, angle=0]{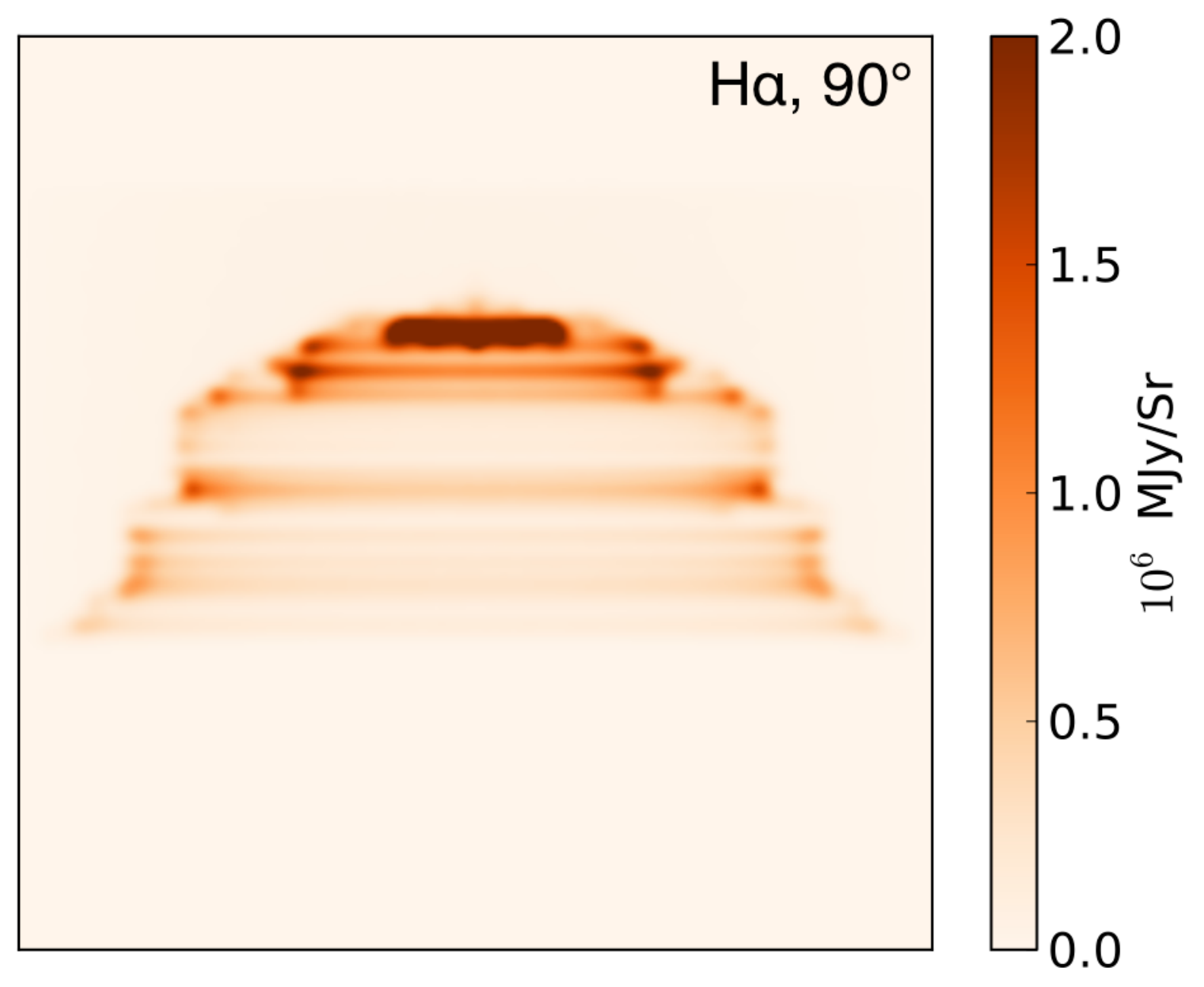}
\includegraphics[scale=0.35,angle=0]{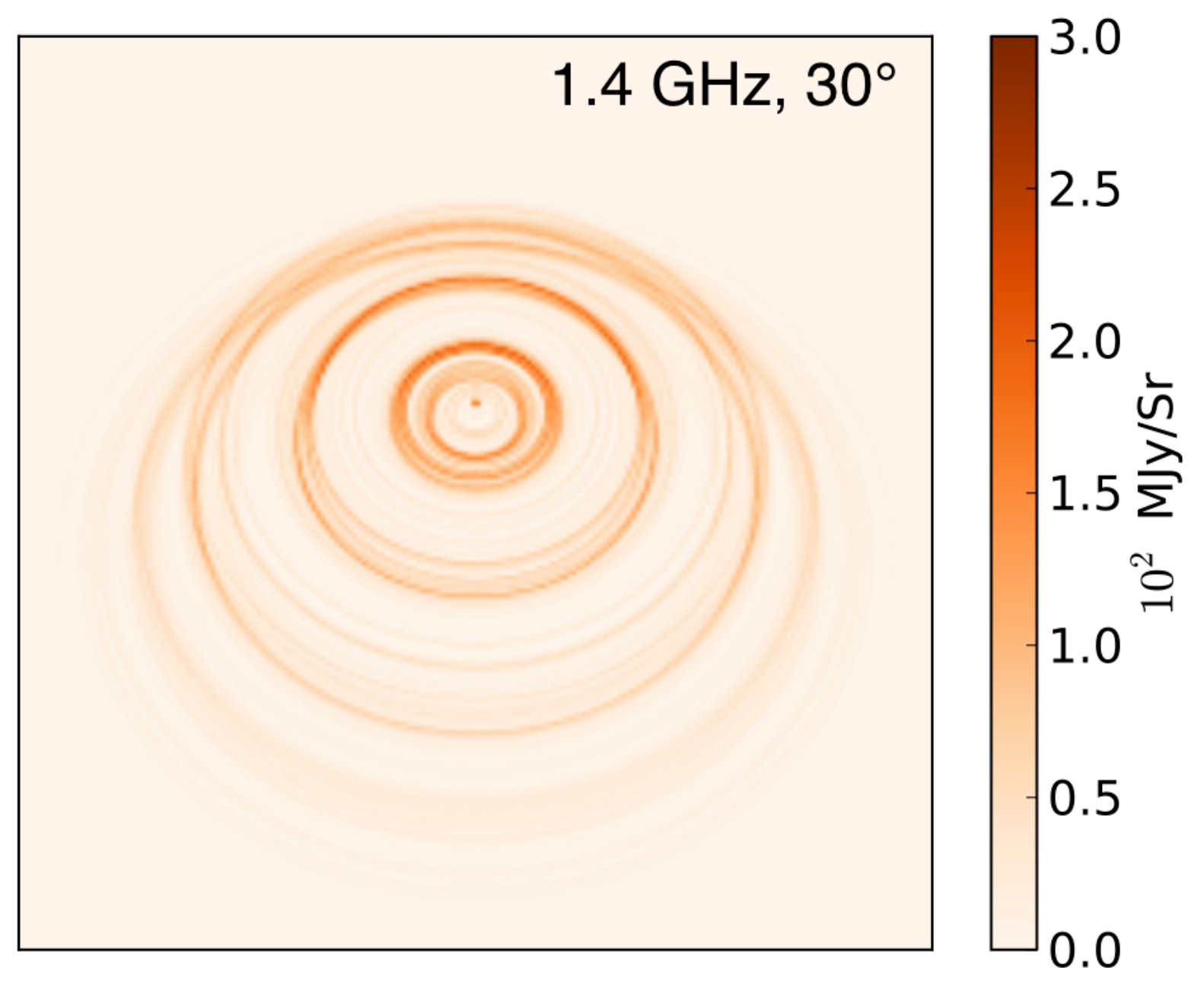}
\includegraphics[scale=0.35,angle=0]{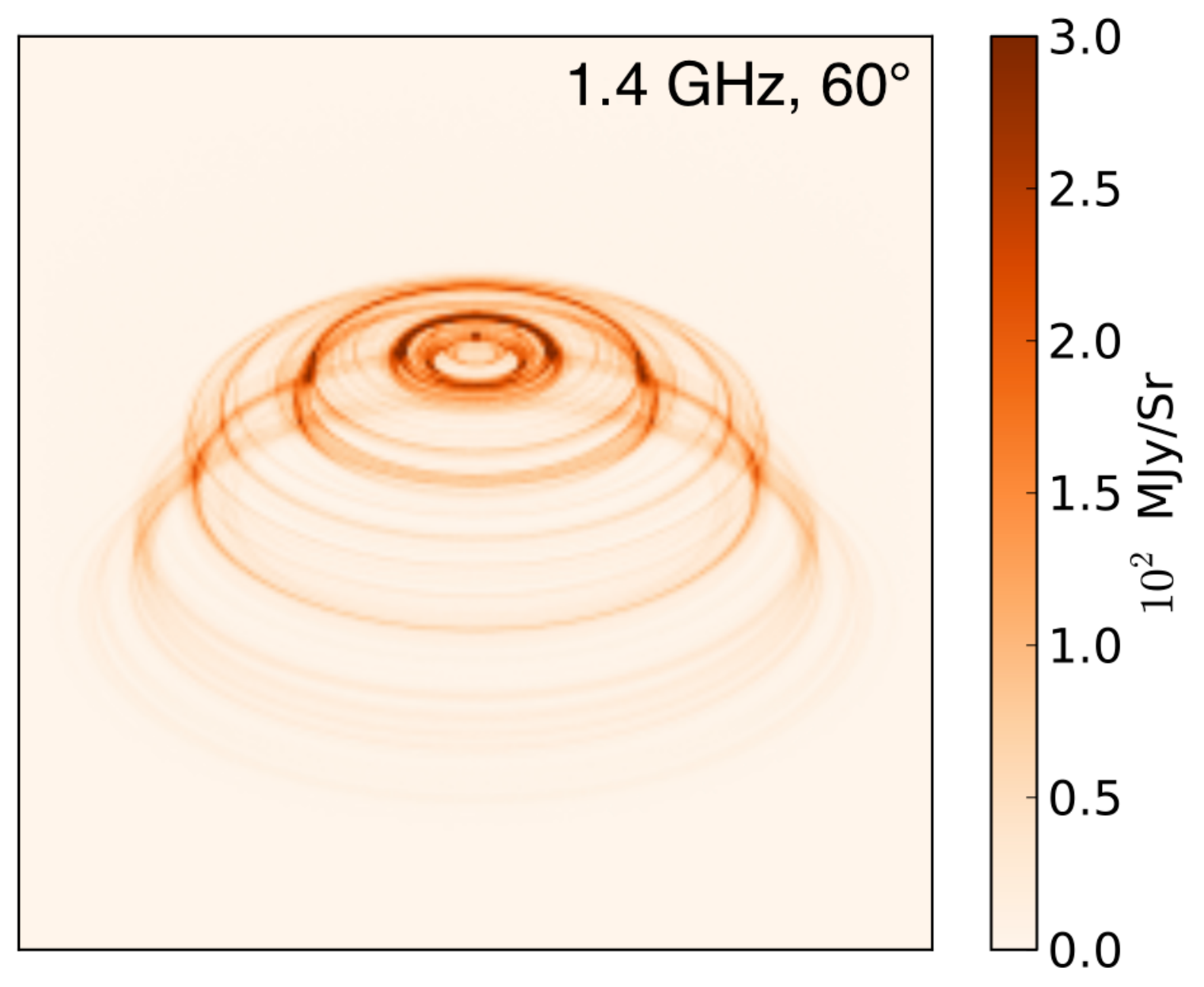}
\includegraphics[scale=0.35,angle=0]{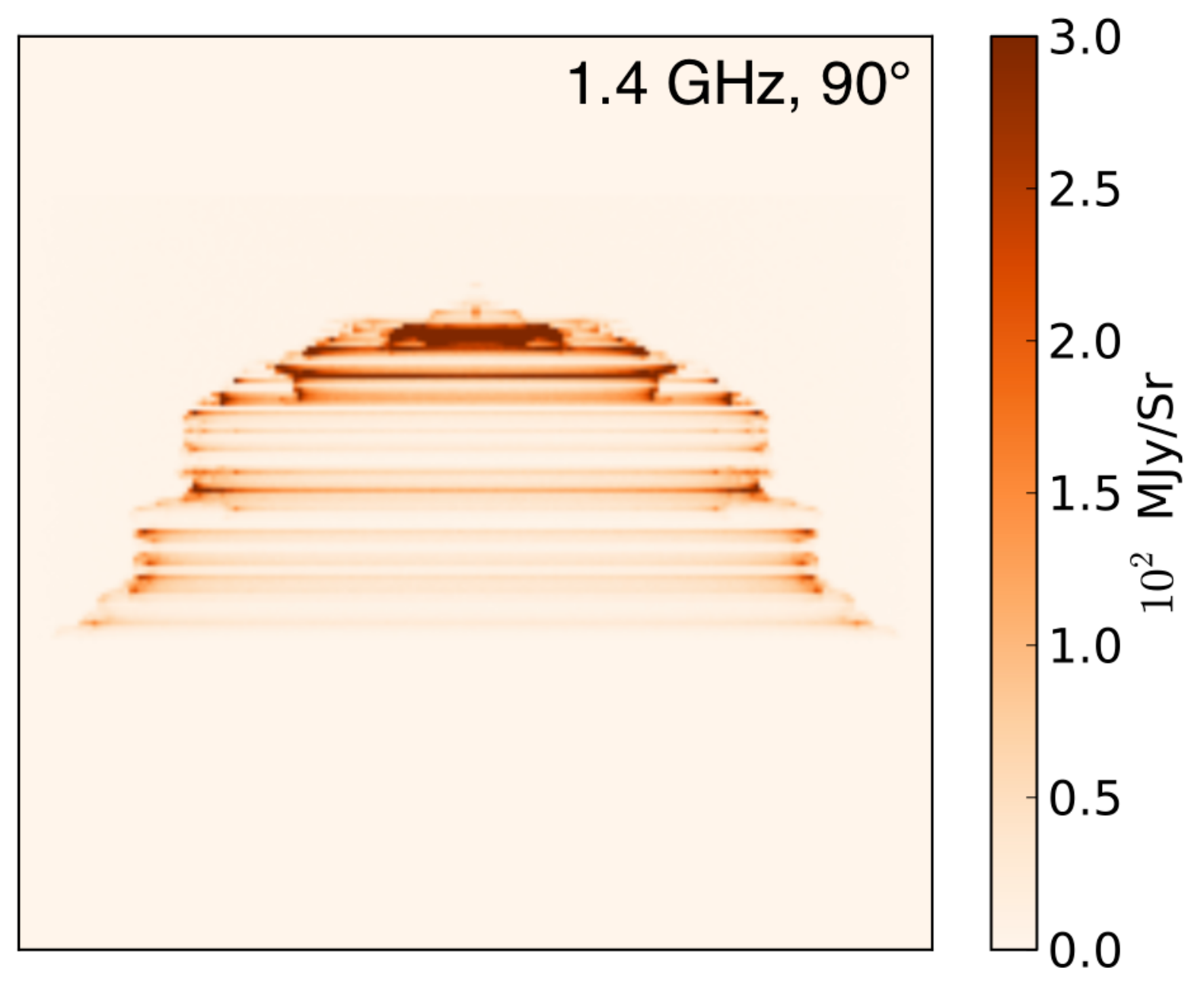}
    \caption{Synthetic images derived from the bow shock
      simulation. The top row and second row show infra-red images at
      $22\mu \rm{m}$ (resolution$=12"$) and $12\mu \rm{m}$ (resolution$=6.5"$) respectively. Third and fourth
      rows show H$\alpha$ (resolution$=1"$) and 1.4~GHz radio emission (no smoothing) respectively.
      Three inclination angles are shown: 30~degrees (left),
      60~degrees (centre), 90~degree (right), where an inclination of
      zero corresponds to a line-of-sight which is parallel to the
      direction of motion of the star. The colour scale shows surface
      brightness in distance independent units of MJy~sr$^{-1}$. The size of the images corresponds to
      2.25~arcmin at a distance of 1~kpc, and the spatial resolution is stated in the individual figure captions.}
\label{fig:synthetic_images}
\end{figure*}

Synthetic images at 22~microns and 12~microns are shown in the first
and second rows of Fig.~\ref{fig:synthetic_images} respectively. The
size of the images corresponds to 2.25~arcmin at a distance of 1~kpc
and the colour scale is in distance independent units of MJy~sr$^{-1}$. To
represent the effect of spatial resolution the 22~micron images have
been smoothed with a 12~arcsec Gaussian filter and the 12~micron
images have been smoothed with a 6.5~arcsec Gaussian filter. These
resolutions have been chosen to match the resolution of WISE at these
wavelengths. The pixel size of 2.75~arcsec was also chosen to match
the properties of WISE.
Images were also generated at 4.6 and 3.4 microns, but
did not produce significant emission from the bow shock. We note that
\cite{Povich_2008} find their observed bow shocks are faint in the
3.4~micron IRAC band, consistent with our lack of emission at this
wavelength. 

At 0 degree inclination (not shown) the shock is viewed face-on and
the emission is is symmetric about the position of the star at the
centre of the image. As we are viewing a 2D model along the axis of
symmetry this is exactly as expected. At 30~degree inclination there
is a noticeable asymmetry to the emission, with a peak ahead of the
star. Although the morphology is not that of a classic bow shock there is still a
clear infra-red excess ahead of the star. At 60 and 90 degree
inclinations a more clearly identifiable bow shock morphology is seen,
with a bright crescent of emission ahead of the star. Higher
inclination angles have higher peak emission due to increased limb
brightening of the shock cone (see \cite{MacLow1991} for further
discussion of the effects of viewing angle). The bow shock morphology
is more clearly seen at 12~microns than at 22~microns as the spatial
resolution is higher (6.5~arcsec compared to 12~arcsec). 
Infra-red observations of bow shocks
show a crescent-shaped structure ahead of the star
frequently with a bright spot near
the apex \citep{Kobulnicky_2010,Peri_2012}. These structures are similar to our 
synthetic observations at 60~degree and 90~degree inclinations.

\subsubsection{SEDs}

\begin{figure}
   \centering
   \includegraphics[scale=0.32,angle=-90]{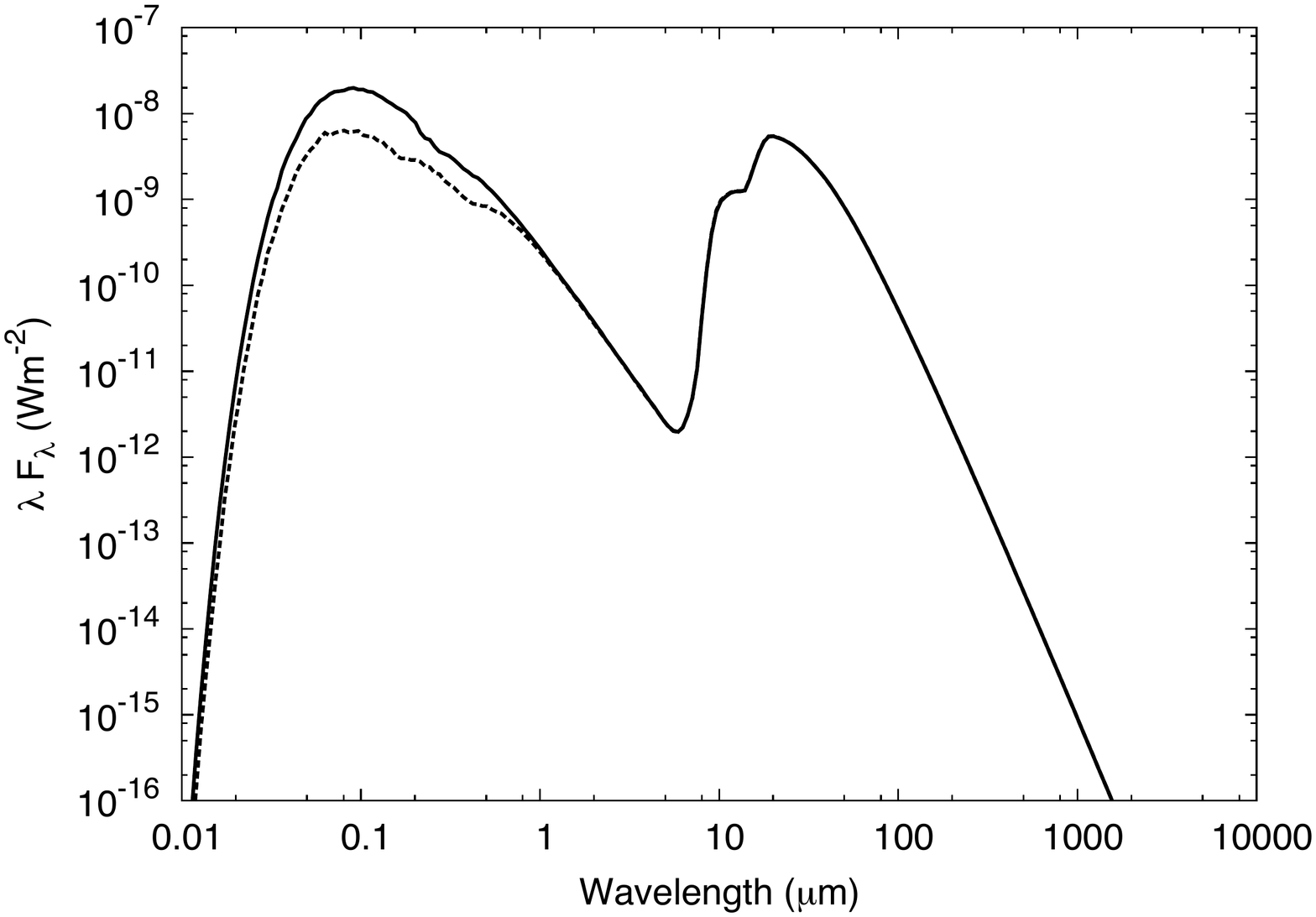}\label{fig:sed_lambdaFlambda}
   \includegraphics[scale=0.32,angle=-90]{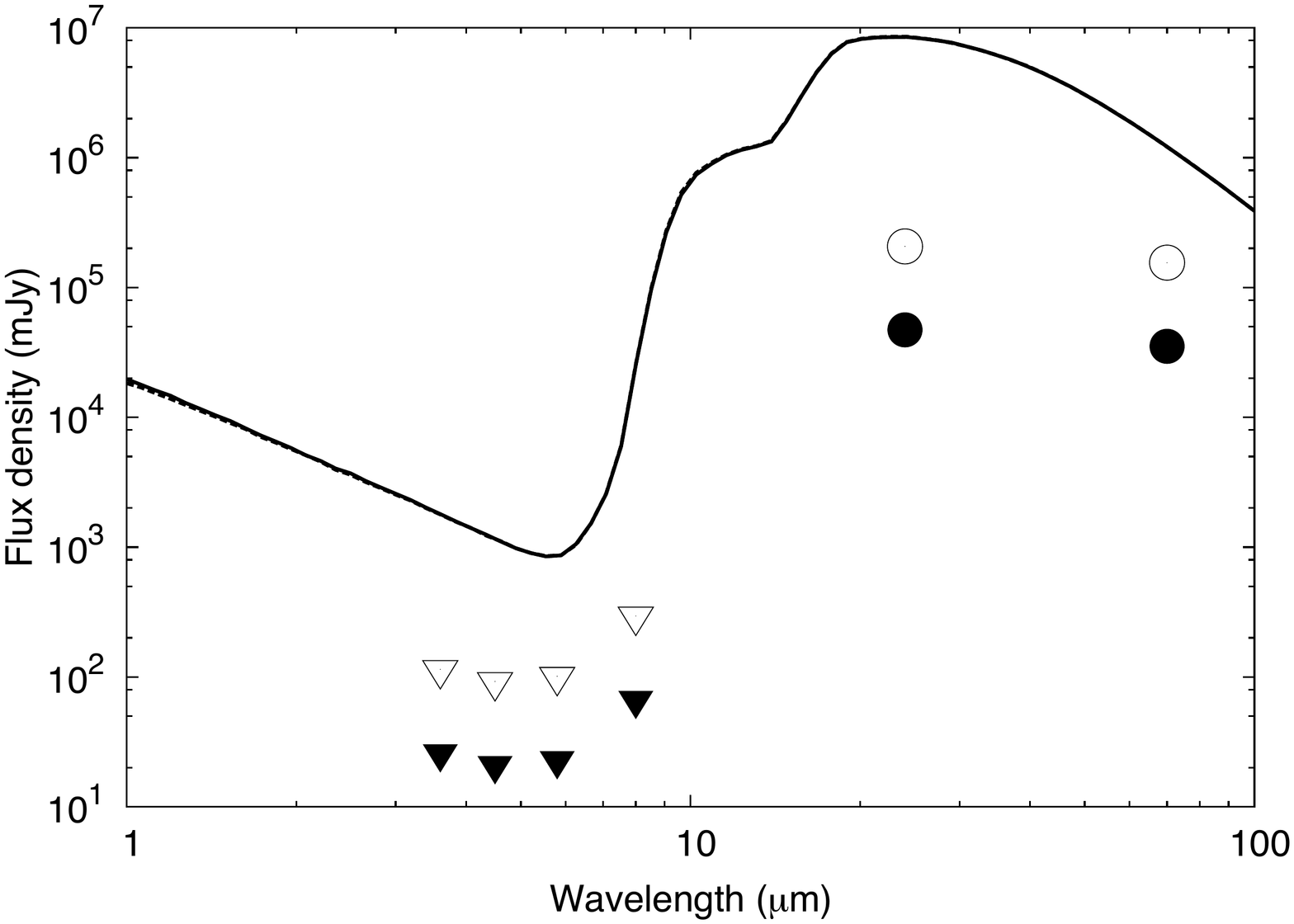}\label{fig:sed_jansky}
   \caption{Spectral energy distributions from the star and bow shock
     viewed at 0~degree inclination (face on; dashed line) and
     90~degree inclination (side on; solid line). 
     Top: SED in SI units plotted as $\lambda\rm{F}_{\lambda}$ vs wavelength
   in microns, bottom: SED in mJy vs wavelength
   in microns.
     At wavelengths
     shorter than 5~$\mu$m the emission is dominated by the star and
     at wavelengths longer than 5~$\mu$m emission from hot dust in the
     bow shock dominates. Solid circles are photometric points from 
     Kobulnicky et al (2010) and solid triangles show upper limits. Open 
     circles and open triangles show the Kobulnicky et al points scaled by a factor of 4.4
     to account for our more luminous star.}
   \label{fig:SEDs}
 \end{figure}

Spectral energy distributions, as seen at 0~degree inclination (face-on; solid line) and
90~degree inclination (side on; dashed line) are shown in
Fig.~\ref{fig:SEDs}. The assumed distance is again 1~kpc.
 The top figure plots the SEDs as $\lambda\rm{F}_{\lambda}$ and shows
 that at wavelengths shorter than 5~$\mu$m the emission is dominated
 by the star, and at wavelengths longer than 5~$\mu$m emission from
 hot dust in the bow shock dominates. This is consistent with the
 infrared images which show bow shock emission at 12 and 22~$\mu$m,
 but not at 3.4 or 4.6~$\mu$m.  At 90~degree inclination the star is
 unobscured by bow shock material and the SED is similar to that of a
 star with an accretion disc viewed face-on (the optical depth along the path from the star to the observer is 0.427 at $0.1~\mu m$). 
 At 0~degree inclination
 bow shock material obscures the star and the SED appears more like an
 SED from a star-disc system where the disc partially blocks direct
 stellar radiation
 (the optical depth along the path from the star to the observer is 2.57 at $0.1~\mu m$).
 The total luminosity of the SED is $1.22\times10^{6}~L_{\odot}$, compared to 
 the stellar luminosity of $L=1.41\times10^{6} L_{\odot}$.

 The lower figure plots the SEDs in mJy over the range 1--100$\mu$m
 and solid points are from Candidate~7 of the \cite{Kobulnicky_2010}) sample, 
 where circles are detections and triangles are upper limits. 
 Candidate~7 is the most massive and energetic star in their
 sample and consequently is the best comparison to our
 model. In Fig.~\ref{fig:sed_jansky} 
 the assumed distance is changed to 2.1~kpc
 to match the distance to candidate~7 in the \cite{Kobulnicky_2010}
 sample.
 \cite{Kobulnicky_2010} find their SED is consistent with a
 temperature of 40~000K for the star and \cite{Hanson2003} finds an
 effective temperature of 46~131.8K and a luminosity $\log \left( L /
   L_{\odot}\right) =5.51$ for the same star. By comparison our model
 star has an effective temperature of 40~000K and $\log \left( L /
   L_{\odot}\right) =6.15$, so is about 4.4 times more luminous but
 with a similar temperature\footnote{The higher luminosity of our
   model star is required in order to be consistent with the stellar
   wind parameters used in Section~\ref{section:vh1_hydro}}. 
   Open 
     circles and open triangles show the \cite{Kobulnicky_2010} Candidate~7 
     points scaled by a factor of 4.4
     to account for our more luminous star.
   Compared
 to the observed photometric data points we have higher levels of
 emission, more than the factor of 4.4 in the stellar
 luminosity, although we have not included extinction effects in our model SED.  
 The peak in the SED due to the bow shock
 appears at shorter wavelengths in our model SED than in the SED
 fitted to Candidate~7 by \cite{Kobulnicky_2010}. However the model has a smaller shock
 stand-off distance ($\sim0.1$pc for the model compared to a projected
 distance of 0.53~pc for candidate~7) so it is reasonable to expect
 higher temperature material in the model bow shock (hence a shorter
 wavelength peak in the SED) and for more stellar radiation to be
 reprocessed through the bow shock (hence higher levels of
 emission). \cite{Meyer2014} find that the bow 
 shock luminosity scales strongly with stand-off distance in their models 
 (stronger that $R_0^3$ according to their fig.~24). Such a strong dependance on $R_0$
 indicates that a quantitative comparison between observed and model 
 SEDs requires tuning the model parameters to produce a similar 
 shock stand off distance to the observed object, as well as the inclusion 
 of extinction. 

\subsection{H$\alpha$ images}

When calculating H$\alpha$ images the emissivity of a grid cell is the
sum of the recombination line emissivity and the dust emissivity (the
latter is described in Section~\ref{section:ir_images}). For this
model the dust emissivity from the bow shock is negligible in
comparison to the recombination emissivity (the emission at $0.66\mu$m
seen in the SED shown in Fig.~\ref{fig:SEDs} is from the star and not
the dust). However scattering from dust is significant at this wavelength (see Fig.~\ref{fig:opacity}), 
affecting emission in bright regions of the synthetic image by several per cent,
so dust opacity is included in the calculations. 

The $H\alpha$ line intensities are calculated using values from \cite{Hummer1987}, 
scaled according to the temperature dependence of the recombination 
coefficient in Table~1 of \cite{Storey1995}. 
The \cite{Hummer1987} line intensities are
calculated relative to the $H\beta$ line intensity given by
\begin{equation}
\epsilon_{H\beta}=  1.235\times10^{-25 }n_e n_{H^+} \left( \frac{T}{10^4\rm{K}} \right) ^{-0.87}
\end{equation}
where the constant term is the emission measure given in
\cite{Hummer1987}. The fiducial values are the case B values 
from \cite{Hummer1987} for a density of $n=10^2~\rm{cm}^{-3}$ and a
temperature of $10^4$~K.

All the gas in our model is highly ionised so 
the H$\alpha$
emissivity is effectively a function of temperature (to
the power -0.87) and density squared.
The emissivity of the stellar wind region is low due to the low density and high temperature. 
H$\alpha$
emission is expected to originate predominantly from the dense
material at the forward shock. This region is bright in
both infra-red emission and H$\alpha$ emission as it is dense, ionised, and contains dust.

H$\alpha$ images are shown in the third row of
Fig.~\ref{fig:synthetic_images}. The images have been blurred with a
1~arcsec Gaussian filter, representative of a high spatial resolution
H$\alpha$ survey (e.g. IPHASS). The elliptical bands of emission seen
at 30 and 60~degree inclinations, and the stripes seen at 90~
inclination, are due to viewing a 2D simulation rotated about the axis
of symmetry. The same effect was seen in the synthetic images of
\cite{Raga1997} shown in their figure~5.  In reality, and in a 3D
simulation, we would expect these features to show bright knots of
emission associated with instabilities at the forward shock (see
Fig.~\ref{fig:torus_photoion_calc}).

The bow shock is more clearly identifiable in the H$\alpha$ images,
with a more prominent bow morphology, due to the higher spatial
resolution. If the H$\alpha$ images are smoothed to the same
resolution as the infra-red images then the bow shock morphology is
very similar. This is unsurprising given that the infra-red
and H$\alpha$ emission both originate from the same region.

\subsection{Radio images}

When generating radio images we calculate thermal free-free emission
from ionised hydrogen and the emissivity of a grid cell is given by
\begin{equation}
\epsilon_{\nu} = n_e^2 \alpha_{kk}\left( \nu,T \right) \exp \left( \frac{-h \nu}{kT}
\right) \frac{2 h \nu^3}{c^2}
\end{equation}
where $\alpha_{kk}\left( \nu,T \right)$ is the free-free absorption
coefficient for hydrogen at a frequency $\nu$ in a cell with
temperature $T$ \citep{MihalasAndMihalas} and $n_e^2$ is the electron
density
The dependance of the emissivity on the square of the electron density
means that dense, ionised material within the forward shock again
dominates the emission.

Synthetic radio observations at 1.4GHz are shown in the fourth row of
Fig.~\ref{fig:synthetic_images}. These images have not been spatially
smoothed and hence show the sharpest features. Bands of emission
associated with the 2D geometry are again seen, which in reality would
correspond to knots of emission at the forward shock. 

\subsection{Emission slices and shock stand-off distance}

\begin{figure*}
   \centering
   \includegraphics[scale=0.3]{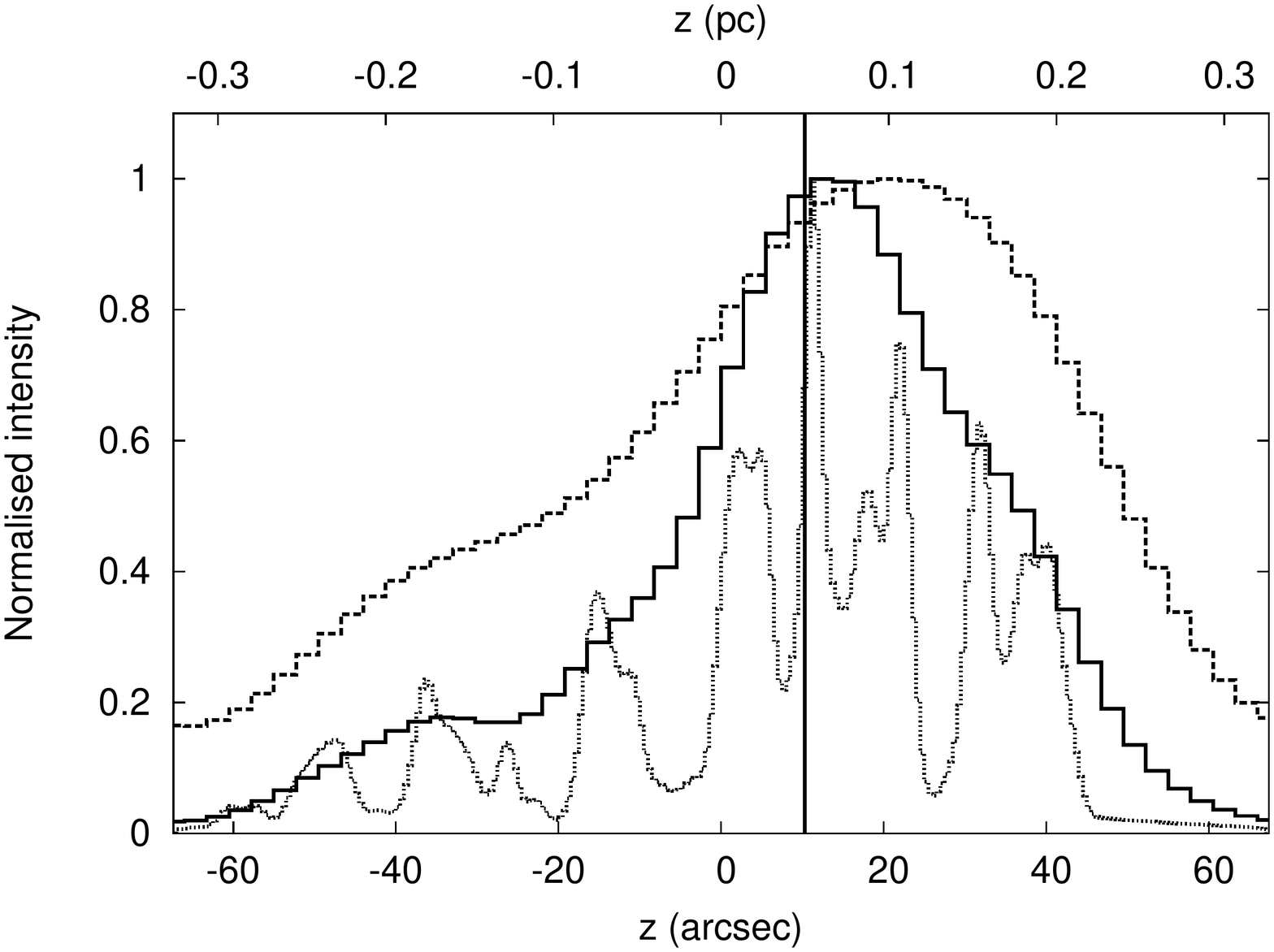}
   \includegraphics[scale=0.3]{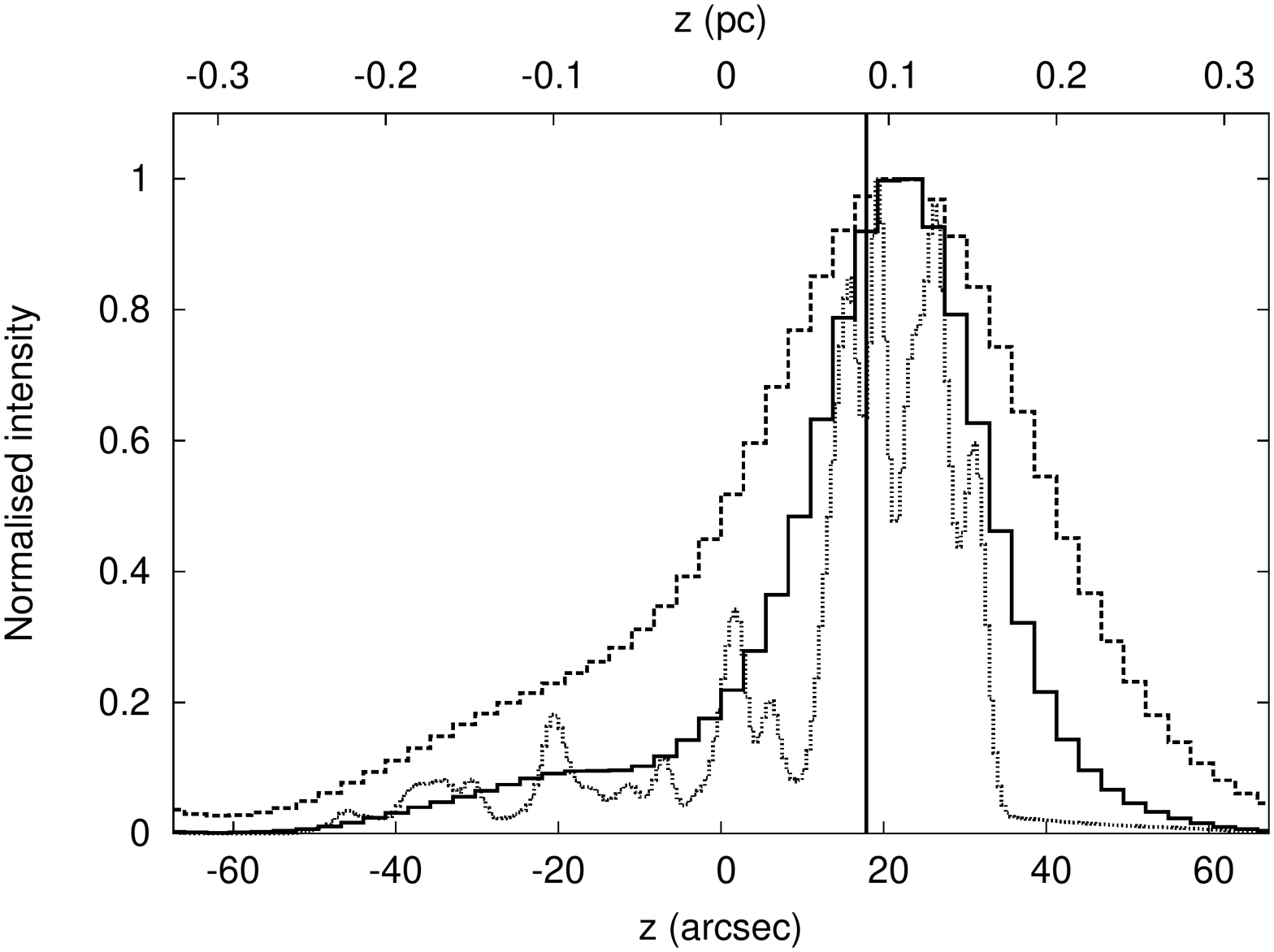}
   \includegraphics[scale=0.3]{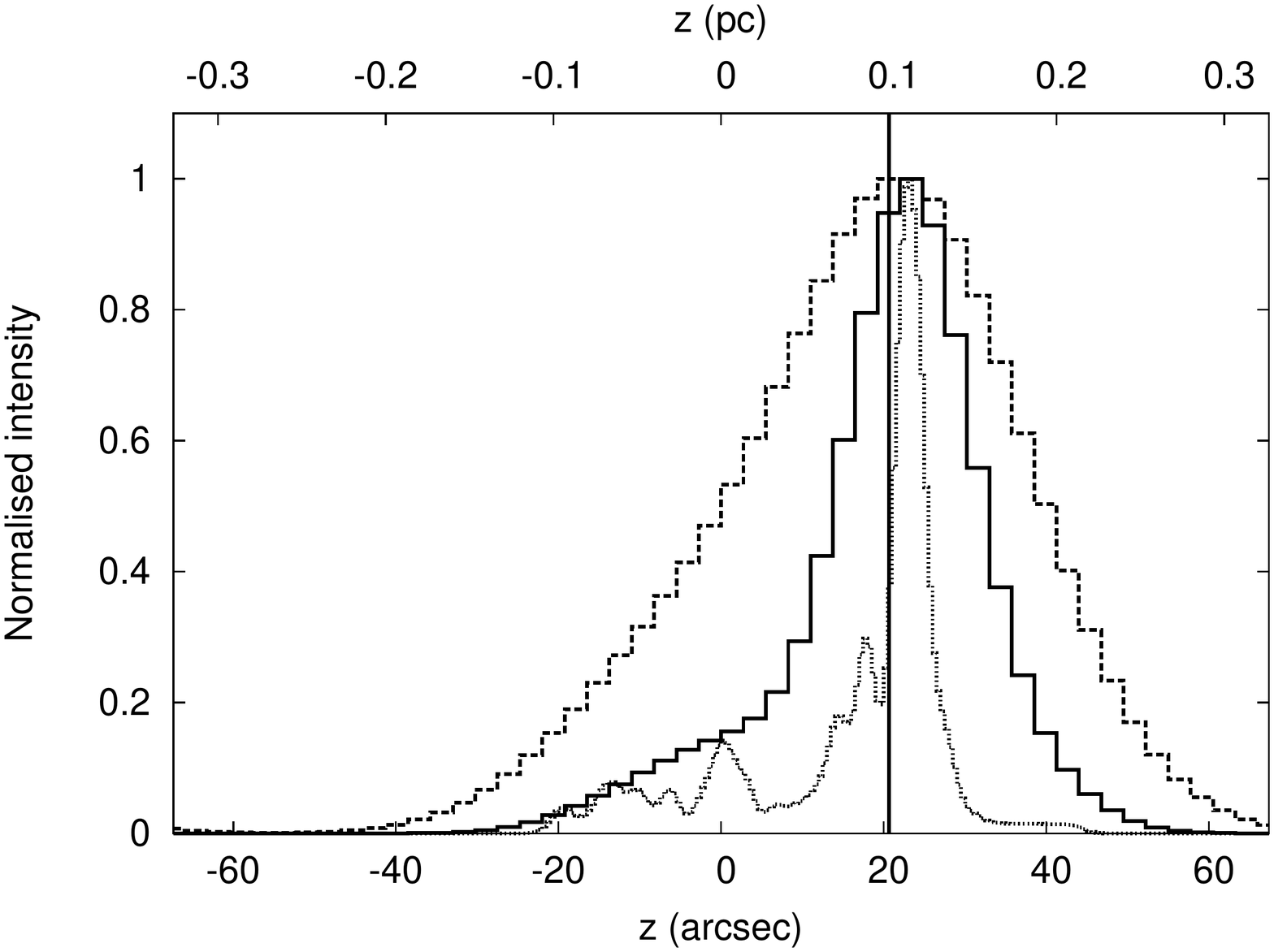}
   \includegraphics[scale=0.3]{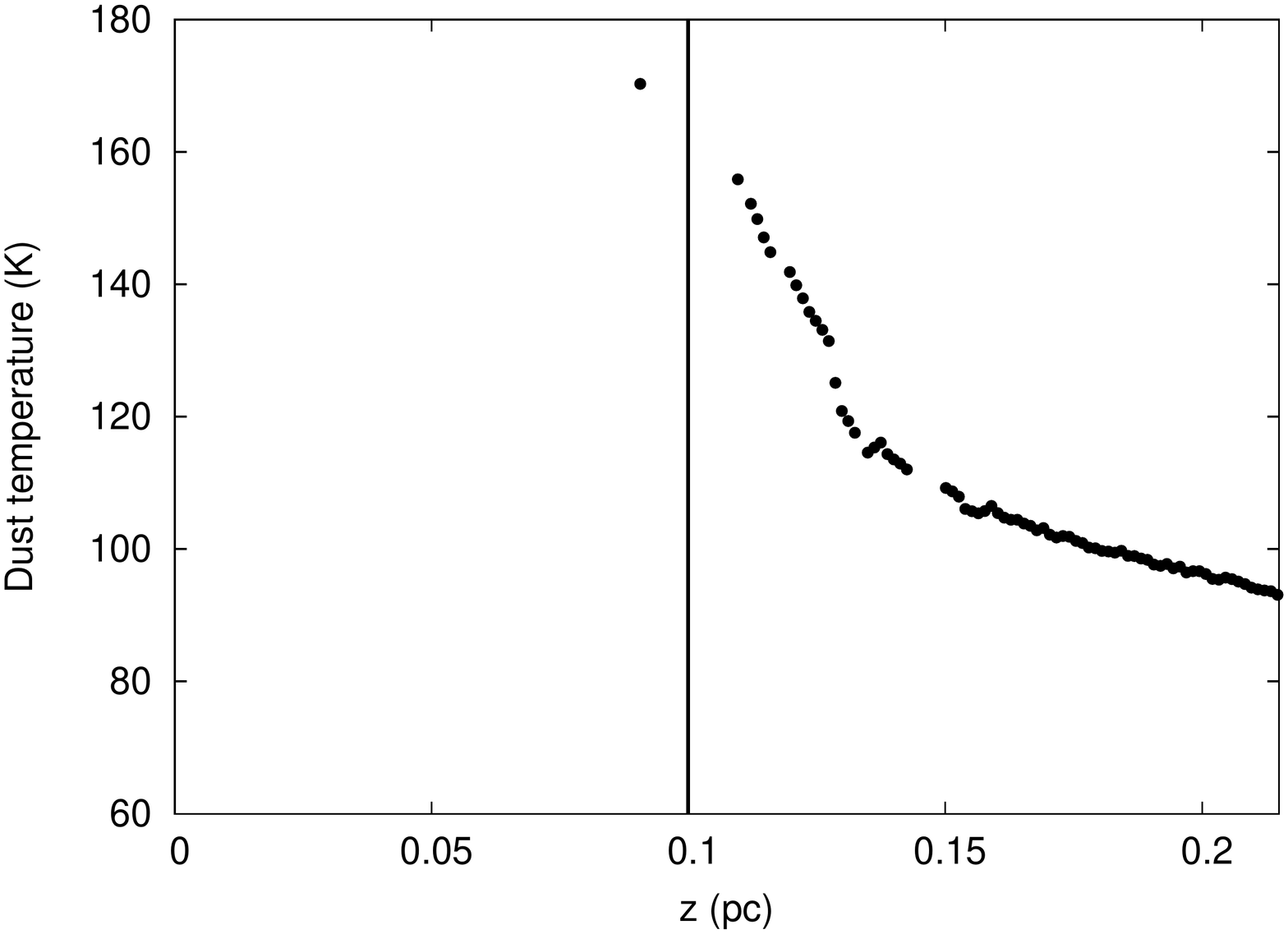}
   \caption{The first three panels show intensity slices along the z-axis at 12$\mu$m (solid line), 22$\mu$m
     (dashed line) and H$\alpha$ (dotted line), for inclinations of 30 (top left), 60 (top right) and 90 (bottom left) degrees.
     For each slice the intensity has been normalised by dividing by the
     maximum intensity. The standoff distance $R_0$, projected by the inclination angle, is shown as a solid vertical line. The bottom right panel shows the dust temperature along the z-axis. The assumed distance is 1~kpc and the star located is at z=0.}
   \label{fig:slices}
 \end{figure*}

Slices of intensity are shown in Fig~\ref{fig:slices} for 12$\mu$m
(solid line), 22$\mu$m (dashed line) and H$\alpha$ (dotted line)
for 30, 60 and 90~degree inclinations. For each slice the intensity has been
normalised, by dividing by the maximum intensity, to enable the shape
of the profiles to be more easily compared. The standoff distance $R_0$, 
projected by the inclination angle, is shown as a solid vertical line.
 The x-axis is shown in parsecs and also arcseconds at a distance of 1~kpc. 
 
\begin{figure}
   \centering
	\includegraphics[scale=0.3]{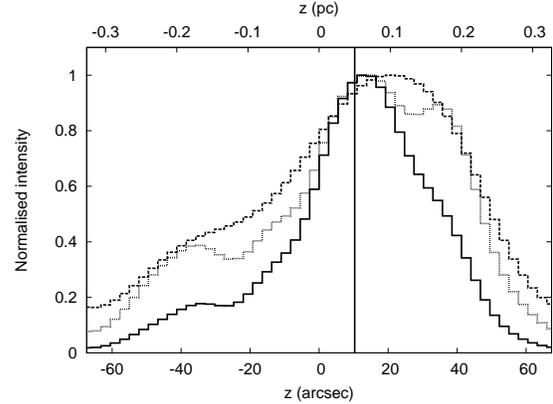}
	\caption{Intensity slices along the z-axis at 12$\mu$m (solid line) and 22$\mu$m
     (dashed line) for a 30~degree inclination angle. The dotted line is the intensity slice from the 22$\mu$m synthetic image smoothed to the same resolution as the 12$\mu$m synthetic image (i.e. 6.5~arcsec) to show resolution effects. The standoff distance $R_0$, projected by the inclination angle, is shown as a solid vertical line.}
	\label{fig:slices_resolution}
\end{figure}

 At 90~degree inclination the peak of the 22$\mu m$ emission is very close to the calculated stand-off distance of 0.1~pc. For the 12$\mu$ and H$\alpha$ emission the peaks are at a slightly greater distance but are also very close to 0.1~pc. The peaks of emission are closer to the calculated stand-off distance than might be expected given the position of the Wilkin analytical solution relative to the forward shock (see Fig.~\ref{fig:torus_photoion_calc}). 
 At 60~degree inclination the infra-red peaks are in approximately the same location as at 90~degree inclination but 
 the H$\alpha$ slice no longer shows a clear maximum.
 At 30~degree inclination the infra-red peaks are clearly separated 
 with the 12$\mu$m peak closer to the star. Figure~\ref{fig:slices_resolution} again shows the 12$\mu$m (solid line) and 22$\mu$m (dashed line) intensity slices (previously plotted in the top left panel of Fig.~\ref{fig:slices}) but also shows a 22$\mu$m slice smoothed to the same resolution as the 12$\mu$m slice (dotted line). This shows that the differences in the 12$\mu$m  and 22$\mu$m profiles are not just due to resolution. The z-axis temperature profile in the region ahead of the star is shown in the bottom right panel of Fig.~\ref{fig:slices} (the star is at $z=0$). 
 A temperature gradient is seen with temperature decreasing in the upstream direction, consistent with the 12$\mu$m emission peaking closer to the star than the 22$\mu$m emission.

\section{Discussion}
\label{section:discussion}

\subsection{Comparison with previous synthetic observations}

Synthetic H$\alpha$ and dust continuum maps of a runaway star have
been previously published by \cite{Raga1997}, also using 2D
hydrodynamical and radiative transfer calculations. Compared to our
simulations \cite{Raga1997} have a much lower ISM density
($1~\rm{cm}^{-3}$ compared to $1000~\rm{cm}^{-3}$). 
In reality the ISM is far from homogeneous and a runaway star will
 encounter a range of ISM properties along its
trajectory, hence both these regimes are of interest. \cite{Raga1997} also
see the highest H$\alpha$ emissivity from the forward shock (see their
fig.~4), and their H$\alpha$ and dust continuum maps (see their
fig.~5) show significant similarities (although they note that the
dust continuum has more extended emission in the bow shock wings).

Synthetic radio images were presented by \cite{MacLow1991} and
compared to observations of ultra-compact H{\sc{II}} regions. In their
model the emission emanates from a limb-brightened ionised shell,
giving a similar morphology to our synthetic radio images. However
their model does not form dense knots of material (the result 
of instabilities in the forward shock/contact discontinuity region) which 
give rise to the smaller scale structures seen in our synthetic observations. 

More recently \cite{Meyer2014}, in their models of main sequence stars, find that ``most of the emission by radiative cooling comes from shocked ISM gas which cools as the gas is advected from the forward shock to the contact discontinuity''. 
In our model the contact discontinuity and forward shock are not separated so there will not be advection from the forward shock to the contact discontinuity. However in both our model and the models of \cite{Meyer2014} most infra-red emission is from dense regions associated with the forward shock.

\subsection{Future model developments}
\label{section:uncertainties}

Although our synthetic observations show an encouraging similarity in
morphology to observations, and are consistent with previous synthetic observations, 
there are aspects of the model which will require further development if all relevant physical processes 
are to be taken into account.

The location and composition of the dust are subject to significant uncertainties but could have a significant influence on observables 
in the infra-red regime. Dust does not necessarily follow the gas distribution in a stellar bow shock \citep{vanMarle2011} with larger dust grains penetrating into the un-shocked stellar wind, and even the smaller grains can reach the wind termination shock \citep{vanMarle2014a}. Variations in dust composition, and presence of PAHs, can have a significant impact on infra-red observables \citep{Kobulnicky_2010,Pavlyuchenkov2013}. Variations in composition can be driven by a number of mechanisms in addition to the effects described by \cite{vanMarle2014a} e.g. differing destruction time scales and the effects of radiation pressure \citep{Ochsendorf2014}.

Although our model provides a thorough treatment of radiative transfer in the gas and dust, and accounts for the effects of shock heating, 
it does not self-consistently couple the hydrodynamics and radiative transfer. It is considerably more computationally expensive to run such a coupled calculation using Monte-Carlo radiative transfer, however such calculations have already been shown to be tractable in 3D using the {\sc{torus}} code \citep{Haworth2012a}. Our current hydrodynamical model also does not include magnetic fields. \cite{vanMarle2014b} examine the effects of magnetic fields on the bow shock around an AGB star. They conclude that the main effect is suppression of instabilities, rather than an influence on the larger scale morphology. As our model involves significantly higher speeds for both the star and the stellar wind we will have a much higher ram pressure from both these components. Consequently the influence of the ISM magnetic field on the large scale structure will be even smaller than that seen by \cite{vanMarle2014b}. However if magnetic fields are effective at suppressing instabilities this will affect the formation of the dense knots of material seen in our model, which are associated with prominent observable features. 

\section{Conclusions}
\label{section:conclusions}

We have generated synthetic images in infrared, H$\alpha$ and radio
regimes for a runaway O star passing though a high density ISM. In all
cases the emission is dominated by knots of dense material formed by
instabilities at the forward shock. This region contains dust but is
also ionised so it is seen not only at infra-red wavelengths (due to
emission from warm dust) but is also seen in the H$\alpha$
recombination line and free-free radio emission. Differences in bow shock morphology are 
largely due to differing spatial resolution and the fundamental morphology is very similar (particularly as we 
have assumed that the dust distribution follows the gas distribution outside the stellar wind).

Synthetic infra-red SEDs show a similar shape to observed bow shock SEDs
\citep{Kobulnicky_2010} but with higher levels of
emission. This can be attributed partly to our model star being
brighter than the observed stars but also to a smaller shock stand off
distance. Achieving a better quantitative match to specific observations would require selecting 
model parameters which better represent the target of the observation. Variations in dust properties and 
the presence of PAHs may also need to be taken into account in order to achieve a quantitative match with observations. 

\section*{Acknowledgments}

We would like to thank Pierre Lesaffre for a helpful review which significantly improved the paper. 
The {\sc{torus}} calculations for this paper were performed on the DiRAC Complexity machine, jointly funded by STFC and the Large Facilities Capital Fund of BIS, and the University of Exeter Supercomputer, a DiRAC Facility jointly funded by STFC, the Large Facilities Capital Fund of BIS, and the University of Exeter.

\bibliographystyle{mn2e}
\bibliography{bowshocks} 

\end{document}